\documentclass[11pt,onecolumn]{article}
\usepackage{url}
\usepackage[colorlinks = true, linkcolor = blue, urlcolor = blue, citecolor = blue, anchorcolor = blue]{hyperref}
\usepackage{amsmath}
\usepackage{amsfonts}
\usepackage{amssymb} 
\usepackage{amsthm}
\usepackage{floatrow}
\usepackage{graphicx}
\usepackage{epstopdf}
\usepackage{subfig}
\usepackage{caption}
\usepackage{cite}
\usepackage{color}
\usepackage{tikz}
\usepackage[ruled,vlined]{algorithm2e}
\usepackage{multicol}
\usepackage{pgfplots}
\tikzset{>=stealth}
\newcommand{\ud}{\,\mathrm{d}}
\newcommand{\sgn}{\text{sgn}}
\usepackage[paper=a4paper,left=16mm,right=15mm,top=14mm,bottom=16mm]{geometry}
\title{Studies on Generalized Fourier Representations and Phase Transforms}
\author{%Pushpendra Singh$^{*}$ and Anubha Gupta$^{\$}$\\\footnote{Author's E-mail address: \texttt{spushp@gmail.com} (P. Singh); \texttt{pushpendrasingh@iitkalumni.org}}
Pushpendra Singh$^{*}$\\\footnote{Author's E-mail address: \texttt{spushp@gmail.com}; \texttt{pushpendrasingh@iitkalumni.org}}	
{\normalsize $^{}$School of Engineering \& Applied Sciences, Bennett University -- Greater Noida, India}
}
\providecommand{\keywords}[1]{\textbf{\textit{Keywords:}} #1}
\date{}
\begin{document}
\maketitle

\begin{abstract} % PRSA 200 words
Fourier representation (FR) is an indispensable mathematical formulation for modeling and analysis of physical phenomenon, engineering systems and signals in numerous applications. In this study, we present the generalized Fourier representation (GFR) that is completely based on the FR of a signal, and introduce the phase transform (PT) which is a special case of the GFR and a true generalization of the Hilbert transform. We derive the PT kernel to obtain any constant phase shift, discuss the various properties of the PT, and demonstrate that (i) a constant phase shift in a signal corresponds to variable time-delays in all harmonics, (ii) to obtain a constant time-delay in a signal, one need to provide variable phase shift in all harmonics, (iii) a constant phase shift is same as the constant time-delay only for single frequency sinusoid. The time derivative and time integral, including fractional order, of a signal can be obtained using the GFR. We propose to use discrete cosine transform (DCT) based implementation to avoid end artifacts due to discontinuities present in both end of the signal. We introduce fractional delay of a discrete time signal using the FR, and present the fast Fourier transform (FFT) implementation of all the above proposed representations.
Using the analytic wavelet transform (AWT), we propose wavelet phase transform (WPT) to obtain a desired phase-shift in a signal under-analysis, and propose the two representations of wavelet quadrature transform (WQT) which is special case of the WPT where phase-shift is $\pi/2$ radians.   
\end{abstract}

\section{INTRODUCTION}
The Fourier representation (FR) of a signal is the most important mathematical formulation for modeling and analysis of physical phenomena and engineering systems. It has been used to obtain solution of problems in almost all fields of mathematics, science, engineering and technologies. FR is the fundamentals of a signal processing, analysis,  information extraction and interpretation. There are many variants of the FR such as continuous-time Fourier series (FS),  Fourier transfrom (FT), Fourier sine transform (FST) and Fourier cosine transform (FCT), discrete-time Fourier transform (DTFT), discrete time Fourier series (DTFS), discrete Fourier transform (DFT), discrete sine transform (DST) and discrete cosine transform (DCT) \cite{IEEECompt,DCTBook}. All these are orthogonal transforms which can be computed by the Cooley--Tukey fast Fourier transform (FFT) algorithms \cite{CT}. Recently, many studies \cite{FQT,rslc8,rslc9,rslc10,rslc11,rslc13} have been performed using the Fourier theory and many applications including signal decomposition and time-frequency analysis of a nonlinear and nonstationary time-series are proposed.   

The DCT was proposed in the seminal paper \cite{IEEECompt} for applications to image processing based pattern recognition and Wiener filtering. The modified DCT (MDCT) \cite{MDCT1} is based on the DCT of overlapping data which uses the concept of time-domain aliasing cancellation \cite{TDAC}.   
The DCT and MDCT are widely-used due to decorrelation and energy compaction properties in many application like image (e.g. JPEG),  video (e.g. Motion JPEG, MPEG, Daala, digital video, Theora) and audio (e.g. MP3, WMA, AC-3, AAC, Vorbis, ATRAC) compression, electrocardiogram data analysis \cite{IEEETNSRE}, and for numerical solution of partial differential equations by spectral methods. There are 8-types of DCTs and 8-types of DSTs depending upon the symmetry about a data point and boundary conditions. 

The Fourier theory based quadrature method was proposed by Gabor \cite{DGabor} in 1946 as a practical approach for obtaining the Hilbert Transform (HT), and thus Gabor analytic signal (GAS) representation of a signal. The GAS has been extensively used in communications engineering, physics, time-frequency-energy (TFE) representation, and signal analysis. The TFE representation of a signal is obtained using the concept of instantaneous frequency (IF) \cite{th19,LCohen,Gabor,Shekel,DEVakman,blBB,th20,th4,th411,LoTa,PSBL,rslc4,rslc5,rslc6,rslc7,rslc12,SSandoval} which is an important parameter in many applications. Recently, using 8-types of DCTs and 8-types of DTSs, 16-types of quadrature Fourier transforms (QFTs) and corresponding Fourier-Singh analytic signal (FSAS) representations are introduced in \cite{FQT} for nonlinear and non-stationary time-series analysis. The FQTs and FSAS representations are alternatives to the HT and GAS representation, respectively. The HT and FQTs are $\pi/2$ phase shifter. However, there is no general method to provide a desired phase shift to signal under analysis. In this study, we propose phase transform (PT), which is based on the proposed generalized Fourier representation (GFR) of a signal, to obtain the desired phase shift and time-delay. We also discuss the various special cases of the GFR namely Fourier representation, PT, time-delay including fractional delay of discrete time signals, time derivative and integral including fractional order, amplitude modulation (AM) and frequency modulation (FM).  

Wavelet transform (WT) uses a wavelet function to analyze signals in various applications and when analyzing wavelet is analytic then corresponding WT is known as analytic wavelet transform (AWT) \cite{awt1,awt2,awt3}. An analytic signal representation of a real-valued and finite energy signal using the AWT is presented in \cite{asbw1,asbw2,asbw3}, where authors obtained the wavelet analytic signal (WAS) using an analytic wavelet function (AWF) with its real part being an even function (Theorem 2 of \cite{asbw1}) and thus imaginary part of AWF is an odd function. The advantages of the WAS in both precision and antinoise performance are also demonstrated.
In this study, we eliminate this restriction (that real part of an AWF has to be an even function to obtain the WAS) and propose the two representations of a wavelet quadrature transform (WQT) and corresponding WAS using any analytic wavelet and corresponding AWT. We also propose two representations of wavelet phase transform (WPT) and show that proposed WQT is a special case of the WPT where phase-shift is $\pi/2$ radians. 

The main contributions of this study are summarized as follows:
\begin{enumerate}

\item Introduction of the generalized Fourier representation (GFR) which is completely based on the Fourier representation of a signal. 

\item Introduction of the phase transform (PT) which is a special case of the GFR, and a true generalization of the Hilbert transform (HT). Using the proposed PT, the desired parse-shift and time-delay can be introduced to a signal under analysis. We derive PT kernel to obtain any constant phase-shift, discuss the various properties of the PT, and demonstrate that the HT is a special case of PT when phase-shift is $\pi/2$ radian. We also provide an extension of the one-denominational PT for two-dimensional image signals in Appendix \ref{MDPSPT}, which can easily be extended for multidimensional signal.
  
\item Using the PT, we demonstrate that (i) a constant phase shift (e.g. HT as $\pi/2$ phase shift) in a signal corresponds to variable time-delays in all harmonics, (ii) to obtain a constant time-delay in a signal, one need to provide variable phase shift in all harmonics, (iii) a constant phase shift is same as the constant time-delay only for single frequency sinusoid. 

\item The time derivative and time integral, including fractional order, of a signal can be obtained using the GFR. We proposed to use discrete cosine transform (DCT) based implementation to avoid end artifacts due to discontinuities present in both end of the signal.  

\item Introduction of the fractional delay of a discrete time signal using the Fourier representations, i.e. DFT, DSTs and DCTs.

\item We present the FFT implementation of all the above proposed representations.  

\item Using the analytic wavelet transform (AWT), we introduce wavelet phase transform (WPT) to obtain a desired phase-shift in a signal under-analysis, and propose the two representations of wavelet quadrature transform (WQT) which is special case of the WPT where phase-shift is $\pi/2$ radians.      
\end{enumerate}

This study is organized as follows: A Generalized Fourier Representation using Fourier series and its various special cases are presented in Section~\ref{meth}. PT using Fourier transform is presented in Section~\ref{meth2}. PT using Fourier sine and cosine transforms is presented in Section~\ref{CTFQT}. The WPT and WQT using AWT are presented in Section~\ref{PT_AWS}. Implementation of the GFR using DFT and DCT is presented in Section \ref{DFT_GFR} and Section \ref{DCT_GFR}, respectively. Simulation results and discussions are presented in Section~\ref{simre}. Section~\ref{con} presents conclusion of the work.

\section{The Generalized Fourier Representation}\label{meth}
In this section we propose the generalized Fourier representation (GFR) and consider its various special cases.   
\subsection{The GFR using Fourier series representation}\label{meth1}
Let $x_{\text{\tiny T}}(t)$ be a real valued periodic signal (i.e., $x_{\text{\tiny T}}(t+T)=x_{\text{\tiny T}}(t), \forall t$) which follows the Dirichlet conditions. The Fourier series expansion of $x_{\text{\tiny T}}(t)$ is given by
\begin{equation}
x_{\text{\tiny T}}(t)=a_0+\sum_{k=1}^{\infty} [a_k\cos(k\omega_0 t)+b_k\sin(k\omega_0 t)], \quad \omega_0=\frac{2\pi}{T}=2\pi f_0 \text{ rad/s, } \label{eq1}
\end{equation}
where $a_0=\frac{1}{T}\int_{t_1}^{t_1+T} x_{\text{\tiny T}}(t)\ud t$, $a_k=\frac{2}{T}\int_{t_1}^{t_1+T} x_{\text{\tiny T}}(t)\cos(k\omega_0 t)\ud t$ and $b_k=\frac{2}{T}\int_{t_1}^{t_1+T} x_{\text{\tiny T}}(t)\sin(k\omega_0 t)\ud t$. 
Using $a_k=r_k\cos(\phi_k)$, $b_k=-r_k\sin(\phi_k)$, where $r_k=\sqrt{a^2_k+b^2_k}$ and $\phi_k=\tan^{-1}({-b_k/a_k})$, one can write
\begin{equation}
x_{\text{\tiny T}}(t)=a_0+\sum_{k=1}^{\infty} [r_k\cos(k\omega_0 t+\phi_k)]. \label{eq2}
\end{equation}

Using Fourier series representation \eqref{eq2}, we hereby propose the GFR as
\begin{equation}
x_{\text{\tiny T}}(t,c_k(t),\alpha_k(t))=a_0 c_0(t) \cos(\alpha_0(t))+\sum_{k=1}^{\infty} [c_k(t) r_k\cos(k\omega_0 t+\phi_k-\alpha_k(t))], \label{eq3}
\end{equation}
where $c_k(t)$ and $\alpha_k(t)$ (for $k=0,1,2,\dots,\infty$) are introduced as amplitude and phase scaling/modulating functions of frequency ($k$) and possibly time (t) as well.
Now we consider the various cases of the GFR as follows:

\textbf{Case 1:} The GFR \eqref{eq3} is Fourier series representation of a signal when $c_k(t)=1$ and $\alpha_k(t)=0$, for all  $t$ and $k=0,1,2,\dots,\infty$.    

\textbf{Case 2:} Using the GFR \eqref{eq3} with $c_k(t)=1$ and $\alpha_k(t)=\alpha_k$, $\forall t, k$, we hereby propose the phase transform (PT) as
\begin{equation}
x_{\text{\tiny T}}(t,\alpha_k)=a_0 \cos(\alpha_0)+\sum_{k=1}^{\infty} [r_k\cos(k\omega_0 t+\phi_k-\alpha_k)], \label{eq4}
\end{equation} 
where $\alpha_k \in [0,2\pi)$ is the phase shift in $k$-th harmonics.
The Hilbert transform (HT), $\hat{x}_{\text{\tiny T}}(t)$, a special case of the PT \eqref{eq4} where it has constant phase shift of  $90^{\circ}$ (i.e. $\alpha_k=\pi/2$, for all $k$), can be defined as
\begin{equation}
\hat{x}_{\text{\tiny T}}(t)=x_{\text{\tiny T}}(t,\pi/2)=\sum_{k=1}^{\infty} [r_k\sin(k\omega_0 t+\phi_k)]. \label{eq5}
\end{equation} 

\textbf{Case 3:} The time-delay of a signal is defined as
\begin{equation}
x_{\text{\tiny T}}(t-t_k)=a_0+\sum_{k=1}^{\infty} [r_k\cos(k\omega_0 t+\phi_k-k\omega_0 t_k)]. \label{eq6}
\end{equation}
From \eqref{eq4}, \eqref{eq5} and \eqref{eq6}, we observe that \textbf{(a)} HT is a constant phase shift which introduces variable time-delays in all harmonics, i.e., $t_k={\pi}/{(2k\omega_0)}$; \textbf{(b)} to obtain a constant time-delay (say, $t_k=t_d$) in a signal, we need to provide variable phase shift in all harmonics, i.e., $\alpha_k=k\omega_0 t_d$; \textbf{(c)} constant phase shift is same as constant time-delay only for single frequency sinusoid (say $k=1$ and hence $\alpha_1=\omega_0 t_d$); \textbf{(d)} variable phase shift is same as variable time-delay only for a zero mean ($a_0=0$) signal if, $\alpha_k=k\omega_0 t_k$, $\forall k$.            

It is interesting to observe that we presented the phase shift of a constant signal in \eqref{eq4} as, $a_0 \cos(\alpha_0)$, which is valid because \textbf{(i)} for the HT, it is zero, \textbf{(ii)} for phase shift of $\pi$, it is multiplied by minus one, and \textbf{(iii)} there is no change in its value if phase shift is zero. However, time-delay operation in a constant signal [e.g., $a_0$ in \eqref{eq6}] does not change its value.

\textbf{Case 4:} Using \eqref{eq2}, one can obtain $\mu$-th order fractional time derivative of a signal as
\begin{equation}
D^{\mu}\{x_{\text{\tiny T}}(t)\}=\frac{a_0t^{-\mu}}{\Gamma \left( 1-\mu \right)}+\sum_{k=1}^{\infty} (k\omega_0)^{\mu} [r_k\cos(k\omega_0 t+\phi_k+\mu\pi/2)], \quad \mu \ge 0, \label{eq7}
\end{equation}
where $\Gamma \left( 1-\mu \right)$ is gamma function.  
The GFR \eqref{eq3} is $\mu$-th order time derivative of the signal when $c_0(t)=\frac{t^{-\mu}}{\Gamma \left( 1-\mu \right)}$, $c_k(t)=(k\omega_0)^{\mu}$, $\alpha_0(t)=0$ and $\alpha_k(t)=-\mu\pi/2$, for all  $t$ and $k$. %Even fractional order derivative can be computed as 

\textbf{Case 5:} Using \eqref{eq2}, one can obtain $\nu$-th order fractional time integral of a signal as
\begin{equation}
D^{-\nu}\{x_{\text{\tiny T}}(t)\}=\frac{a_0t^{\nu}}{\Gamma \left( 1+\nu \right)}+\sum_{k=1}^{\infty} (k\omega_0)^{-\nu} [r_k\cos(k\omega_0 t+\phi_k-\nu\pi/2)], \quad \nu \ge 0. \label{eq8}
\end{equation}
The GFR \eqref{eq3} is $\nu$-th order time integral of the signal when $c_0(t)=\frac{t^{\nu}}{\Gamma \left( 1+\nu \right)}$, $c_k(t)=(k\omega_0)^{-\nu}$, $\alpha_0(t)=0$ and $\alpha_k(t)=\nu\pi/2$, for all  $t$ and $k$. 

\textbf{Case 6:} Using \eqref{eq2}, one can obtain amplitude modulated (AM) signal for one arbitrary but fixed value of $k$ (say $k=1$) and thus carrier frequency $\omega_c=\omega_0$, $\alpha_1(t)=0$, $c_0(t)=0$, $c_1(t)=(A_m+m(t)) \ge 0$, where $m(t)$ is a message signal whose maximum frequency $\omega_m<<\omega_c$.    

\textbf{Case 7:} Using \eqref{eq2}, one can obtain angle modulated, frequency modulated (FM) and phase modulated (PM), signal for one arbitrary but fixed value of $k$ (say $k=1$) and thus carrier frequency $\omega_c=\omega_0$, $c_0(t)=0$, $c_k(t)=1$, $\alpha_1(t)=m(t)$, where $m(t)$ is a message signal whose maximum frequency $\omega_m<<\omega_c$.    

%\section{Phase Transform using Fourier transform, Fourier sine and cosine transform}\label{meth20}
\section{Phase Transforms using the FT, FCT and FST}\label{meth20}
In this section, to obtain a desired phase shift in a signal, we present PT using Fourier transform (FT), Fourier cosine transform (FCT), and Fourier sine transform (FST), along with FFT implementation of various cases of the proposed GFR.   
\subsection{PT using Fourier Transform}\label{meth2}
The Fourier transform (FT) and inverse FT (IFT) pairs of a signal are defined as
\begin{equation}
\begin{aligned}
X(\omega)&=\int_{-\infty}^{\infty} x(t) \exp(-j\omega t) \ud t, \quad -\infty < \omega < \infty, \\
x(t)&={\frac{1}{2\pi}}\int_{-\infty}^{\infty}  X(\omega) \exp(j\omega t) \ud \omega, \quad -\infty < t < \infty,
\end{aligned}\label{FT1}
\end{equation}
subject to the existence of the integrals, and these pairs can be denoted by $x(t)\rightleftharpoons X(f)$, where $\omega=2 \pi f$. From the definitions of FT and IFT \eqref{FT1}, one can observe that $X(0)=\int_{-\infty}^{\infty} x(t) \ud t=0$, and $x(0)=\frac{1}{2\pi}\int_{-\infty}^{\infty} X(\omega) \ud \omega=0$, provided that $x(t)$ and $X(\omega)$ are zero-mean functions, respectively. Moreover, an odd function is a zero-mean function, however, reverse is not always true.

First, we consider subtle details of the \textit{zero-mean} function $x_1(t)=\exp(-a|t|)\,\sgn(t)$, where $a>0$ and sign function is defined as \cite{SH}:
\begin{equation}
\sgn(t) =
\begin{cases}
1,  \quad t>0, \\
0, \quad t = 0,\\
-1   \quad t<0.
\end{cases}\label{FTexa1}
\end{equation} 
The FT of $x_1(t)$ is evaluated as \cite{SH}: $X_1(\omega)=\frac{-j2\omega}{a^2+\omega^2}$, which implies that $X_1(f)=0$ at $f=0, \forall a>0$. It is pertinent to notice that, $\lim_{a\to0} X_1(f)=\frac{1}{j\pi f}$, $\lim_{f\to0^{-}}\frac{1}{j\pi f}=j\infty$, and $\lim_{f\to0^{+}}\frac{1}{j\pi f}=-j\infty$ which implies $\lim_{f\to0}\frac{1}{j\pi f}$ does not exist, and therefore $\lim_{a\to0} X_1(0)$ is not defined, however, the FT of a \textit{zero-mean} function $x_1(t)$ constrains that $\lim_{a\to0} X_1(0)=0$.   
Thus, we write the FT $S(f)$ of the \textit{zero-mean} sign function, $s(t)=\sgn(t)=\lim_{a\to0} x_1(t)$, as
\begin{equation}
\lim_{a\to0} X_1(f)=S(f) =
\begin{cases}
0,  \quad f=0, \\
\frac{1}{j\pi f}, \quad f \ne 0.
\end{cases}\label{FTexa3}
\end{equation}
Using the duality principle of the FT, i.e. $X(t)\rightleftharpoons x(-f)$, one can obtain, $jS(t)\rightleftharpoons j\,\sgn(-f) =-j\,\sgn(f)$, as sign function is an odd function. We denote $jS(t)= h(t)$, and thus obtain
\begin{equation}
H(f) =-j\,\sgn(f), \text{ and } h(t) =
\begin{cases}
0,  \quad t=0, \\
\frac{1}{\pi t}, \quad t \ne 0.
\end{cases} \label{FTexa2}
\end{equation} 
Now, we compute $h(t)$ from $H(f)$ using IFT \eqref{FT1} as $h(t)={\frac{-j}{2\pi}} \left[ \int_{-\infty}^{0} - \exp(j\omega t) \ud \omega+\int_{0}^{\infty}  \exp(j\omega t) \ud \omega\right]$ and obtain
\begin{equation}
h(t)=\frac{1}{\pi} \int_{0}^{\infty} \sin(\omega t) \ud \omega,
\label{htk}
\end{equation}
and, clearly, $h(0)=0$, which is also required by the definition \eqref{FTexa2}. The function $\frac{1}{\pi t}$ is the well-known HT kernel which is not defined at origin, moreover, $\lim_{t\to0^{-}}\frac{1}{t}=-\infty$, and $\lim_{t\to0^{+}}\frac{1}{t}=\infty$, therefore  $\lim_{t\to0}\frac{1}{t}$ does not exist. Thus, we have presented a trivial but important modification to HT kernel in \eqref{FTexa2} by defining it at origin, and obtained its integral form in \eqref{htk}. These definitions of the HT kernel is further supported by the discrete time HT kernel \cite{rslc9}, defined as $h[n]=\frac{(1-\cos(\pi n))}{\pi n}$, which is also zero at origin (i.e., $h[0]=0$), and it can be obtained by using the discrete counter part of \eqref{htk}, i.e., $h[n]=\frac{1}{\pi} \int_{0}^{\pi} \sin(\Omega n) \ud \Omega$.    
 
For a real-valued signal $x(t)$, we can write Gabor analytic signal (GAS) as 
\begin{equation}
z(t)=x(t)+j\hat{x}(t)={\frac{1}{\pi}}\int_{0}^{\infty}  X(\omega) \exp(j\omega t) \ud \omega, \quad -\infty < t < \infty,
\label{FT2}
\end{equation}
where real part is the original signal and imaginary part is the HT of real part. Thus, we write
\begin{equation}
\begin{aligned}
x(t)&={\frac{1}{\pi}}\int_{0}^{\infty}  |X(\omega)| \cos(\omega t+\phi(\omega)) \ud \omega, \quad -\infty < t < \infty, \\
\hat{x}(t)&={\frac{1}{\pi}}\int_{0}^{\infty}  |X(\omega)| \sin(\omega t+\phi(\omega)) \ud \omega, \quad -\infty < t < \infty,
\end{aligned}\label{FT3}
\end{equation}
where $X(\omega)=X_{\text{i}}(\omega)+jX_{\text{r}}(\omega)=|X(\omega)|\exp(j\phi(\omega))$, $\phi(\omega)=\tan^{-1}(X_{\text{i}}(\omega)/X_{\text{r}}(\omega))$, and the GFR corresponding to \eqref{eq3} as
\begin{equation}
{x}(t,c(\omega,t),\alpha(\omega,t))={\frac{1}{\pi}}\int_{0}^{\infty} c(\omega,t)  |X(\omega)| \cos(\omega t+\phi(\omega)-\alpha(\omega,t)) \ud \omega. \label{GFRFT}
\end{equation}

We hereby define the PT ${x}(t,\alpha(\omega))$ of signal $x(t)$ using the IFT as 
\begin{equation}
{x}(t,\alpha(\omega))={\frac{1}{\pi}}\int_{0}^{\infty}  |X(\omega)| \cos(\omega t+\phi(\omega)-\alpha(\omega)) \ud \omega, \quad -\infty < t < \infty, \label{FT4}
\end{equation}
where $\alpha(\omega)$ is the phase shift introduced in the signal $x(t)$. The PT, ${x}(t,\alpha(\omega))$ defined in \eqref{FT4}, is the real part of the PT of analytic signal hereby defined as
\begin{equation}
z(t,\alpha(\omega))={\frac{1}{\pi}}\int_{0}^{\infty}  X(\omega) \exp(j(\omega t-\alpha(\omega))) \ud \omega,
\label{HTPT}
\end{equation} 
and therefore obtain transfer function (TF) as
\begin{equation}
H(\alpha(\omega)) =
\begin{cases}
e^{-j\alpha(\omega)},  \quad \omega \ge 0, \\
%e^{-j\alpha(0)}, \quad \omega=0, \\
e^{j\alpha(\omega)},  \quad \omega<0.
\end{cases}\label{FSPT_TF}
\end{equation}
For example, to obtain a constant time-delayed signal $x(t-t_0)$ from input signal $x(t)$, we set $\alpha(\omega)=\omega t_0$, and therefore from \eqref{FT4} and \eqref{FT1}, $x(t,\alpha(\omega))=x(t-t_0) \implies {\frac{1}{2\pi}}\int_{-\infty}^{\infty}  X(\omega) \exp(j\omega t) \exp(-j\omega t_0) \ud \omega={\frac{1}{\pi}}\int_{0}^{\infty}  |X(\omega)| \cos(\omega t+\phi(\omega)-\omega t_0) \ud \omega$.

Thus, we have defined a general phase shifter of a signal in \eqref{FT4} which is a generalization of the IFT as well as HT because it is (i) IFT if $\alpha(\omega)=0$, (ii) HT if $\alpha(\omega)=\pi/2$, and (iii) when $\alpha(\omega)=\alpha$, we obtain its impulse response, designated as constant phase-transform (PT) kernel, from \eqref{FT4} as
\begin{equation}
\hbar(t,\alpha)= \cos(\alpha)\delta(t)+\sin(\alpha)h(t) \implies \delta(t,\alpha)= \cos(\alpha)\delta(t)+\sin(\alpha)\delta(t,\pi/2), \label{FS1}
\end{equation}   
where $\delta(t)$ is the Dirac delta function, $h(t)=\hat{\delta}(t)=\delta(t,\pi/2)$ is the impulse response or HT kernel as defined in \eqref{FTexa2} or \eqref{htk}, $\hbar(t,\alpha)=\delta(t,\alpha)$ and $\delta(t,0)=\delta(t)$. 
The direct derivation of the PT kernel defined in \eqref{FS1} from \eqref{FT4} is elementary. Here, we present indirect proof of that as follows: one can easily show [using \eqref{FT1} or \eqref{FT2} and setting $x(t)=\delta(t)\Leftrightarrow X(\omega)=1$ and $\phi(\omega)=0$] that, $\delta(t)={\frac{1}{2\pi}}\int_{-\infty}^{\infty}  \cos(\omega t) \ud \omega={\frac{1}{\pi}}\int_{0}^{\infty}  \cos(\omega t) \ud \omega$, and $h(t)={\frac{1}{\pi}}\int_{0}^{\infty}\sin(\omega t) \ud \omega$.
%We know, $\text{FT}\{h(t)\}=-j\,\text{sgn}(f)$ which also implies $h(t)={\frac{1}{\pi}}\int_{0}^{\infty}\sin(\omega t) \ud \omega$.
Using these facts, from \eqref{FT4} we obtain, $\hbar(t,\alpha)={\frac{1}{\pi}}\int_{0}^{\infty}\cos(\omega t-\alpha) \ud \omega={\frac{1}{\pi}}\int_{0}^{\infty}[\cos(\alpha)\cos(\omega t)+\sin(\alpha)\sin(\omega t)] \ud \omega$, and thus \eqref{FS1}.  

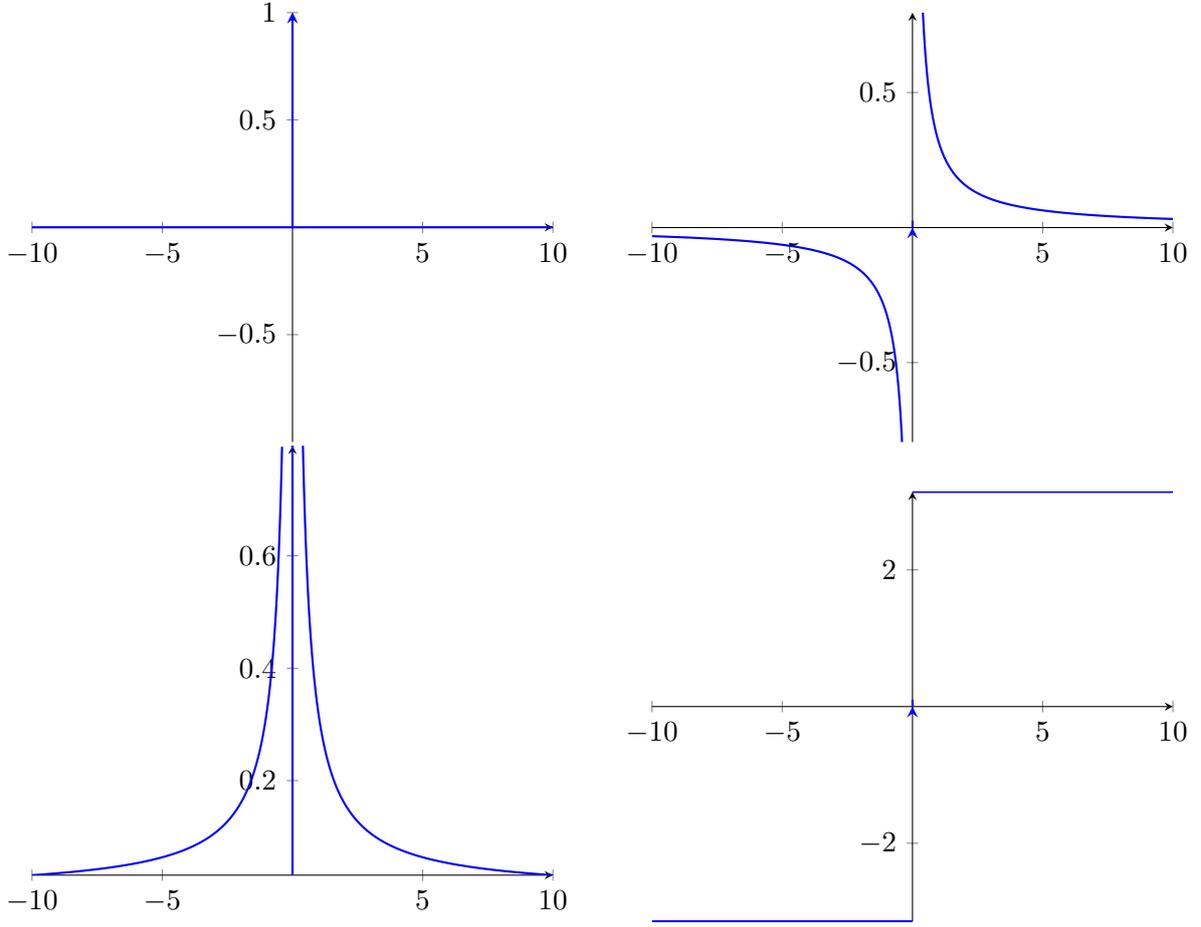
\begin{figure}[!t]
\begin{center}
	\begin{tikzpicture}[scale=1]
	\begin{axis}[axis lines=middle,samples=2000]
	\addplot[blue,thick,domain=-10:10] {0};	
	\draw[->,blue,thick] (axis cs:0,0) -- (axis cs:0,1);
	\end{axis}
	\end{tikzpicture}
	\quad
	\begin{tikzpicture}[scale=1]
	\begin{axis}[axis lines=middle,samples=2000]
	\addplot[blue,thick,domain=-10:-0.4] {1/(x*pi)};
	\addplot[blue,thick,domain=0.4:10] {1/(x*pi)};
	\draw[->,blue,thick] (axis cs:0,0) -- (axis cs:0,0);
	\end{axis}
	\end{tikzpicture}
	\\
\begin{tikzpicture}[scale=1]
\begin{axis}[axis lines=middle,samples=2000]
\addplot[blue,thick,domain=-10:-0.4] {1/abs(x*pi)};
\addplot[blue,thick,domain=0.4:10] {1/abs(x*pi)};
\draw[->,blue,thick] (axis cs:0,0) -- (axis cs:0,1);
\end{axis}
\end{tikzpicture}
\quad
\begin{tikzpicture}[scale=1]
\begin{axis}[axis lines=middle,samples=2000]
\addplot[blue,thick,domain=-10:0] {-pi};
\addplot[blue,thick,domain=0:10] {pi};
\draw[->,blue,thick] (axis cs:0,0) -- (axis cs:0,0);
\end{axis}
\end{tikzpicture}
\end{center}
\caption{Plots of delta function $\delta(t)$ (top-left), HT kernel $h(t)$ (top-right); amplitude $a_{\delta}(t)$ (bottom-left), and phase $\phi_{\delta}(t)$ (bottom-right) of analytic signal kernel \eqref{phad}.} \label{Fig:dkp}
\end{figure}

We can obtain kernel of analytic signal and compute the phase difference between $\delta(t)$ and its HT kernel $h(t)=\delta(t,\pi/2)$ defined in \eqref{FTexa2} as
\begin{equation}
\begin{aligned}
z_{\delta}(t)&=\delta(t)+jh(t) = a_{\delta}(t)e^{j\phi_{\delta}(t)}, \text{ where } \\
a_{\delta}(t)&=\delta(t)+|h(t)|, \text{ and }\\
\phi_{\delta}(t)& = \tan^{-1}\left(\frac{h(t)}{\delta(t)}\right)= \frac{\pi}{2} \, \sgn(t)=
\begin{cases}
\pi/2,  \quad t>0, \\
0, \quad t = 0,\\
-\pi/2,   \quad t<0.
\end{cases}
\end{aligned} \label{phad}
\end{equation}
Figure \ref{Fig:dkp} shows plots of (top-left) delta function $\delta(t)$, (top-right) HT kernel $h(t)$ \eqref{FTexa2}; amplitude $a_{\delta}(t)$ (bottom-left), and phase $\phi_{\delta}(t)$ (bottom-right) of kernel of analytic signal \eqref{phad} which is well-defined for all time including at origin due to the modification presented in HT kernel \eqref{FTexa2} by defining it at origin. We can observe that if HT kernel \eqref{FTexa2} is not defined or has singularity at origin, then phase $\phi_{\delta}(t)$ would be undefined at origin in \eqref{phad}. The kernel of analytic signal can be also written as $z_{\delta}(t)=\delta(t)+jh(t) = {\frac{1}{\pi}}\int_{0}^{\infty}  \exp(j\omega t) \ud \omega$ or $z_{\delta}(t)=\lim_{\sigma \to 0^+} {\frac{1}{\pi}}\int_{0}^{\infty}  \exp(-\omega (\sigma-jt)) \ud \omega=\lim_{\sigma \to 0^+} \frac{1}{\pi}\left(\frac{\sigma}{\sigma^2+t^2}+\frac{jt}{\sigma^2+t^2}\right)$, $\frac{1}{\pi}\int_{-\infty}^{\infty}\frac{\sigma}{\sigma^2+t^2} \ud t=1$, which implies (i) $\delta(t)=\lim_{\sigma \to 0^+}\frac{1}{\pi}\left(\frac{\sigma}{\sigma^2+t^2}\right)$ and $h(t)=0$, for $t=0$, (ii) $\delta(t)=0$ and $h(t)=\frac{1}{\pi t}$, for $t\ne 0$.

The arbitrary but fixed phase shifter defined in \eqref{FS1} is a linear time-invariant (LTI) system model, thus its output can be written as convolution of input with impulse response, i.e., 
\begin{equation}
{x}(t,\alpha)=H_{\alpha}\{x(t)\}=x(t)*\hbar(t,\alpha)=\cos(\alpha)x(t)+\sin(\alpha)\hat{x}(t), \label{FS2}
\end{equation}
or ${x}(t,\alpha)=\cos(\alpha)x(t,0)+\sin(\alpha){x}(t,\pi/2)$, where $x(t,0)=x(t)$ and $x(t,\pi/2)=\hat{x}(t)$ as defined in \eqref{FT3}.
Clearly, the PT of a time-domain signal $x(t)$ is another time-domain and phase shifted signal $x(t,\alpha)$.
There are some obvious properties of the PT \eqref{FS2} which follow directly from the definition such as one can easily show $H_\alpha=\cos(\alpha)I+\sin(\alpha)H$, where $I$ and $H$ are identity and HT operators, respectively, i.e., $I\{x(t)\}=x(t)$ and $H\{x(t)\}=\hat{x}(t)=x(t,\pi/2)$; inverse PT $H^{-1}_{\alpha}=H_{-\alpha}$ (or $H^{-1}_{\alpha}H_{\alpha}=H_{\alpha}H^{-1}_{\alpha}=I$); $H^{m}_{\alpha}=H_{m\alpha}$, $H_{\alpha_2} H_{\alpha_1} =H_{\alpha_1}H_{\alpha_2}=H_{\alpha_1+\alpha_2}$ which implies $H^m_{\alpha}\{x(t)\}=x(t,m\alpha)$, $H_{\alpha_2}\{H_{\alpha_1}\{x(t)\}\}=H_{\alpha_2}\{{x}(t,\alpha_1)\}={x}(t,\alpha_1+\alpha_2)$, and ${x}(t,\alpha_1+\alpha_2)=x(t)$ if $\alpha_2+\alpha_1={2\pi m}, \forall m \in \mathbb{Z}$. Now, we explore some basic properties of the proposed PT \eqref{FS2} as follows:
\begin{enumerate}
\item \textbf{Linearity:} The PT is a linear operator, i.e., $H_{\alpha}\{a_1x_1(t)+a_2x_2(t)\}=a_1x_1(t,\alpha)+a_2x_2(t,\alpha)$ for arbitrary scalars $a_1$ and $a_2$, functions  
$x_1(t)$ and $x_2(t)$.
\item \textbf{The PT of a constant signal:} For any constant $c$, $H_{\alpha}\{c\}=c \cos(\alpha)$.

\item \textbf{Time-shifting and time-dilation:} If $x(t)$ has PT $x(t,\alpha)$, then $x(t-t_0)$ has PT $x(t-t_0,\alpha)$, and $x(at)$ has PT $\cos(\alpha)x(at)+\sin(\alpha) \text{sgn} (a)\hat{x}(at)$ where $a\ne 0$. %sgn(a)ˆg(at) (assuming a 6= 0).

\item \textbf{Relation with the Fourier Transform:}
The Fourier transform of PT kernel is $H(f,\alpha)=\cos(\alpha) + \sin(\alpha)\left(-j\,\text{sgn}(f)\right)$ or
\begin{equation}
H(f,\alpha) =
\begin{cases}
e^{-j\alpha},  \quad f>0, \\
\cos(\alpha), \quad f=0, \\
e^{j\alpha},  \quad f<0,
\end{cases}\label{ftHST}
\end{equation}
thus PT provides $-\alpha$ and $\alpha$ phase shifts to positive and negative frequencies, respectively, and when, $\alpha=\pi/2$, it becomes HT. 
If $x(t)$ has Fourier transform $X(f)$, then, $X(f,\alpha)=X(f) H(f,\alpha)$.
It is to be noted that a generalized Hilbert transform to obtain a phase shift to any angle $\alpha$ is defined in \cite{PHT}, where author defined $H(f,\alpha)=0$ for $f=0$, and for $f\ne 0$, $H(f,\alpha)$ is defined same as \eqref{ftHST}. 

\item \textbf{Orthogonality:} If $x(t)$ is a real-valued energy signal (i.e., $E=\left< x(t),x(t)\right>$), then inner product of $x(t)$ and $x(t,\alpha)$ is given by
\begin{equation}
\left< x(t),x(t,\alpha)\right>=\cos(\alpha) \left< x(t),x(t)\right> + \sin(\alpha) \left< x(t),\hat{x}(t)\right> \implies \cos(\alpha)= \frac{\left< x(t),x(t,\alpha)\right>}{\left< x(t),x(t)\right>}, \label{ortho}
\end{equation}  
as $\left< x(t),\hat{x}(t)\right>=0$, and thus they are orthogonal only for phase shift $\alpha=\frac{\pi}{2}(2m+1), m\in \mathbb{Z}$. 

\item \textbf{Energy:} If $x(t)$ is a real-valued energy signal, then $x(t,\alpha)$ is also a real-valued energy signal and its energy ($E_\alpha$) is computed by inner product of $x(t,\alpha)$ with itself as
\begin{equation}
E_\alpha=\left<x(t,\alpha),x(t,\alpha)\right>=\cos^2(\alpha) \left< x(t),x(t)\right> + \sin^2(\alpha) \left< \hat{x}(t),\hat{x}(t)\right>, \label{enrgy}
\end{equation}  
and for zero mean signal, energy is preserved in HT, i.e., $\left< x(t),x(t)\right> = \left< \hat{x}(t),\hat{x}(t)\right>$, so energy is preserved in the proposed PT. 

\item \textbf{Time-derivative:} The PT of the derivative of a signal is the derivative of the PT, i.e., \\ $H_{\alpha}\{\frac{\ud}{\ud t}x(t)\}=\frac{\ud}{\ud t} H_{\alpha}\{ x(t)\}$.

\item \textbf{PT of product of low-pass and high-pass signal:}\label{PTP} Let $x_1(t)$ be a low-pass signal such that its FT $X_1(f) = 0$ for $|f|>f_0$ and let $x_2(t)$ be a high-pass signal with $X_2(f) = 0$ for $|f|<f_0$. Then, PT $H_{\alpha}\{x_1(t)x_2(t)\}=x_1(t)H_{\alpha}\{x_2(t)\}=x_1(t)x_2(t,\alpha)$. One can show it easily using the property of HT (i.e., Bedrosian theorem \cite{Bedrosian,JLB}) as $H\{x_1(t)x_2(t)\}=x_1(t)\hat{x}_2(t)$. Thus, to obtain the PT of product of a low-pass signal and a high-pass signal, only the high-pass signal needs to be phase shifted.

\item \textbf{PT of an analytic signal:} From \eqref{HTPT} or \eqref{FS2}, we obtain $H_{\alpha}\{ z(t)\}=x(t,\alpha)+j\hat{x}(t,\alpha)=z(t)e^{-j\alpha} =[x(t)+j\hat{x}(t)][{\cos(\alpha)}-j\sin(\alpha)]$ which gives $H_{\alpha}\{ z(t)\}=\cos(\alpha)x(t)+\sin(\alpha)\hat{x}(t)+j[-\sin(\alpha)x(t)+\cos(\alpha)\hat{x}(t)]=x(t,\alpha)+jx(t,\alpha+\pi/2)$. Thus, we can compute $x(t,\alpha)$ by considering the real part of the $H_{\alpha}\{ z(t)\}$.  
\end{enumerate}

\textbf{Observation 3.1:}\label{obj1} Here, we consider three examples of PT property \ref{PTP}, which will be used to obtain a unique PT and thus HT of 2D and higher dimensional signals, as follows. First, we consider a signal with frequencies $\omega_1,\omega_2 \ge 0$ as $x(t)=\cos(\omega_1 t)\cos(\omega_2 t)=\frac{1}{2}[\cos(\omega_1 t+\omega_2 t)+\cos(\omega_1 t-\omega_2 t)]=\frac{1}{2}[\cos(\omega_1 t+\omega_2 t)+\cos(\omega_2 t-\omega_1 t)]$. We obtain the PT of signal $x(t)$ as (i) $x(t,\alpha)=\frac{1}{2}[\cos(\omega_1 t+\omega_2 t-\alpha)+\cos(\omega_1 t-\omega_2 t-\alpha)]=\cos(\omega_1 t-\alpha)\cos(\omega_2 t)$, if $\omega_1>\omega_2$, (ii) $x(t,\alpha)=\frac{1}{2}[\cos(\omega_1 t+\omega_2 t-\alpha)+\cos(\omega_2 t-\omega_1 t-\alpha)]=\cos(\omega_1 t)\cos(\omega_2 t-\alpha)$, if $\omega_2>\omega_1$. Next, we consider $x(t)=\sin(\omega_1 t)\sin(\omega_2 t)$, then we obtain PT as (i) $x(t,\alpha)=\sin(\omega_1 t-\alpha)\sin(\omega_2 t)$, if $\omega_1>\omega_2$, and (ii) $x(t,\alpha)=\sin(\omega_1 t)\sin(\omega_2 t-\alpha)$, if $\omega_2>\omega_1$. Finally, we consider $x(t)=\sin(\omega_1 t)\cos(\omega_2 t)$, then we obtain PT as (i) $x(t,\alpha)=\sin(\omega_1 t-\alpha)\cos(\omega_2 t)$, if $\omega_1>\omega_2$, and (ii) $x(t,\alpha)=\sin(\omega_1 t)\cos(\omega_2 t-\alpha)$, if $\omega_2>\omega_1$. Thus, one can observe that to obtain a unique PT, in the phase argument of sin and cos, before introducing a phase $\alpha$, one must consider $(\omega_1 t-\omega_2 t)$ or $(\omega_2 t-\omega_1 t)$ depending upon whether $\omega_1>\omega_2$ or $\omega_2>\omega_1$.     

\subsection{PT using Fourier sine and cosine transforms}\label{CTFQT}
The Fourier cosine transform (FCT) and inverse FCT (IFCT) pairs, of a signal, are defined as
\begin{equation}
\begin{aligned}
X_c(\omega)&=\sqrt{\frac{2}{\pi}}\int_{0}^{\infty} x(t) \cos(\omega t) \ud t, \quad \omega \ge 0 \\
x(t)&=\sqrt{\frac{2}{\pi}}\int_{0}^{\infty}  X_c(\omega) \cos(\omega t) \ud \omega, \quad t \ge 0,
\end{aligned}\label{fct}
\end{equation}
subject to the existence of the integrals, i.e.,  $x(t)$ is absolutely integrable ($\int_{0}^{\infty} |x(t)| \ud t<\infty$) and its derivative $x'(t)$ is piece-wise continuous in each bounded subinterval of $[0, \infty)$.

The Fourier cosine quadrature transform (FCQT), $\tilde{x}_c(t)$, using the FCT of signal of $x(t)$ is defined in \cite{FQT} as 
\begin{equation}
\tilde{x}_c(t)=\sqrt{\frac{2}{\pi}}\int_{0}^{\infty}  X_c(\omega) \sin(\omega t) \ud \omega, \label{fqtc1}
\end{equation}
\begin{equation}
\tilde{X}_c(\omega)=\sqrt{\frac{2}{\pi}}\int_{0}^{\infty} \tilde{x}_c(t)  \sin(\omega t) \ud t, \label{ifqtc}
\end{equation}
where 
\begin{equation}
\tilde{X}_c(\omega) =
\begin{cases}
0,  \quad \omega=0, \\
{X}_c(\omega),  \quad \omega >0.
\end{cases}\label{ifqtc1}
\end{equation}
The FSAS, using the FCQT, is defined in \cite{FQT} as
\begin{equation}
\tilde{z}_c(t)=x(t)+j\tilde{x}_c(t)=\sqrt{\frac{2}{\pi}}\int_{0}^{\infty}  X_c(\omega) \exp(j\omega t) \ud \omega, \label{fsas1}
\end{equation}
where real part is the original signal and imaginary part is the FQT of real part.

We hereby define the PT ${x}(t,\alpha)$ of signal $x(t)$ using the FCT as 
\begin{equation}
{x}(t,\alpha(\omega))=\sqrt{\frac{2}{\pi}}\int_{0}^{\infty}  X_c(\omega) \cos(\omega t-\alpha(\omega)) \ud \omega, \label{fqtc}
\end{equation}
where $\alpha(\omega)$ is the frequency dependent phase shift. If phase shift is constant or independent of frequency, i.e. $\alpha(\omega)=\alpha$, then we obtain
\begin{equation}
{x}(t,\alpha)=\sqrt{\frac{2}{\pi}}\int_{0}^{\infty}  X_c(\omega) \cos(\omega t-\alpha) \ud \omega=\cos(\alpha)x(t)+\sin(\alpha)\tilde{x}_c(t), \label{fqtc11}
\end{equation} 
where $0\le \alpha < 2\pi$ is the phase shift. 

Thus, we have defined a general phase shifter of a signal which is a generalization of the IFCT as well as FCQT because it is (i) IFCT if $\alpha=0$, and (ii) FCQT if $\alpha=\pi/2$.

The Fourier sine transform (FST) and inverse FST (IFST) pairs, of a signal, are defined as
\begin{equation}
\begin{aligned}
X_s(\omega)&=\sqrt{\frac{2}{\pi}}\int_{0}^{\infty} x(t) \sin(\omega t) \ud t, \\
x(t)&=\sqrt{\frac{2}{\pi}}\int_{0}^{\infty}  X_s(\omega) \sin(\omega t) \ud \omega, 
\end{aligned} \label{fst}	
\end{equation}
subject to the existence of the integrals.	
The Fourier sine quadrature transform (FSQT), $\tilde{x}_s(t)$, using the FST of signal $x(t)$ is defined in \cite{FQT} as 
\begin{equation}
\tilde{x}_s(t)=\sqrt{\frac{2}{\pi}}\int_{0}^{\infty}  X_s(\omega) \cos(\omega t) \ud \omega, \label{fqts}
\end{equation}
\begin{equation}
\tilde{X}_s(\omega)=\sqrt{\frac{2}{\pi}}\int_{0}^{\infty} \tilde{x}_s(t)  \cos(\omega t) \ud t, \label{i1fqtc}
\end{equation}
where one can observe that both representations, defined as FST of $x(t)$ in \eqref{fst} and FCT of $\tilde{x}_s(t)$ in \eqref{i1fqtc}, are same for all frequencies, i.e. ${X}_s(\omega)=\tilde{X}_s(\omega)$.		
The FSAS, using the FSQT, is defined in \cite{FQT} as
\begin{equation}
\tilde{z}_s(t)=\tilde{x}_s(t)+jx(t)=\sqrt{\frac{2}{\pi}}\int_{0}^{\infty}  X_s(\omega) \exp(j\omega t) \ud \omega, \label{fsas2}
\end{equation}
where imaginary part is the original signal and real part is the FQT of imaginary part.

We hereby define the PT ${x}(t,\alpha(\omega))$ of signal $x(t)$ using the FST as 
\begin{equation}
{x}(t,\alpha(\omega))=\sqrt{\frac{2}{\pi}}\int_{0}^{\infty}  X_s(\omega) \sin(\omega t-\alpha(\omega)) \ud \omega. \label{fqtc2}
\end{equation}
If $\alpha(\omega)=\alpha$, then we obtain 
\begin{equation}
{x}(t,\alpha)=\sqrt{\frac{2}{\pi}}\int_{0}^{\infty}  X_s(\omega) \sin(\omega t-\alpha) \ud \omega=\cos(\alpha)x(t)-\sin(\alpha)\tilde{x}_s(t), \label{fqtc22}
\end{equation}
where $0\le \alpha(\omega) < 2\pi$ is the phase shift.  
Thus, we have defined a general phase shifter of a signal which is a generalization of the IFST as well as FSQT because it is (i) IFST if $\alpha=0$, and (ii) FSQT if $\alpha=\pi/2$.

The FQTs, presented in \eqref{fqtc1} and \eqref{fqts}, are different from the HT \eqref{FT3} by definition itself. The proposed FSTP representations, defined in \eqref{fqtc} and \eqref{fqtc2}, are effective phase shifter which can be used in various applications such as envelop detection, IF estimation, time-frequency-energy representation and analysis of nonlinear and nonstationary data.  

%%%%%%%%%%%%%%%%%%%%%%%%%%%%%%%%%%%%%%%%%%%
\subsection{PT using Continuous Analytic Wavelet Transform}\label{PT_AWS}

In this subsection, we use analytic wavelet transform (AWT) to define wavelet phase transform (WPT) of a signal and show that wavelet quadrature transform (WQT) is special case of the WPT when phase-shift is $\pi/2$ radians. The wavelet transform of a signal $x(t)\in L^2(\mathbb{R})$ is defined as \cite{awt1,awt2}
\begin{equation}
%\begin{aligned}
W_\psi(s,\tau)=\langle x(t),\psi_{s,\tau}(t)\rangle=\int_{-\infty}^{\infty}  x(t) \psi^{*}_{s,\tau}(t) \ud t, \label{awt1}
%W_\psi(s,\tau)&=\frac{1}{2\pi}\int_{-\infty}^{\infty} \Psi^{*}(s\omega) X(\omega)e^{j\omega \tau}\ud \omega,
%\end{aligned} 
\end{equation}
which can be represented in the Fourier domain as
\begin{equation}
%\begin{aligned}
%W_\psi(s,\tau)&=\langle x(t),\psi_{s,\tau}(t)\rangle=\int_{-\infty}^{\infty}  x(t) \psi^{*}_{s,\tau}(t) \ud t, \\
W_\psi(s,\tau)=\frac{\sqrt{|s|}}{2\pi}\int_{-\infty}^{\infty} \Psi^{*}(s\omega) X(\omega)e^{j\omega \tau}\ud \omega, \label{awt0}
%\end{aligned}
\end{equation}
where asterisk denotes the complex conjugate operation, $\psi_{s,\tau}(t)=\frac{1}{\sqrt{|s|}}\psi\left(\frac{t-\tau}{s}\right)$ with scaling and translation parameters, $s,\tau \in \mathbb{R}, s\ne 0$, is a family of wavelet daughters, and $\psi(t) \in L^2(\mathbb{R})$ is a mother wavelet function which has finite energy, zero mean and satisfy the admissibility condition \cite{awt3}
\begin{equation}
C_\psi=\int_{-\infty}^{\infty} \frac{|\Psi(\omega)|^2}{|\omega|} \ud \omega < \infty, \label{awt2}
\end{equation}
where $\Psi(\omega)$ is the Fourier transform of the wavelet $\psi(t)$.
The original signal $x(t)$ can be recovered using the inverse wavelet transform defined as
\begin{equation}
x(t)=\frac{1}{C_\psi} \int_{-\infty}^{\infty} \int_{-\infty}^{\infty} W_\psi(s,\tau) \psi_{s,\tau}(t)  \ud \tau \frac{1}{s^2}\ud s. \label{awt31}
\end{equation}
 
If $\psi(t)$ is an analytic wavelet, then $\Psi(\omega)=0$ for $\omega<0$ and \eqref{awt1} is the AWT \cite{awt1,awt2}, moreover \eqref{awt0} and \eqref{awt2} can be re-presented as
\begin{equation}
\begin{aligned}
W_\psi(s,\tau)&=\frac{\sqrt{|s|}}{2\pi}\int_{0}^{\infty} \Psi^{*}(s\omega) X(\omega)e^{j\omega \tau}\ud \omega,\\
C_\psi&=\int_{0}^{\infty} \frac{|\Psi(\omega)|^2}{|\omega|} \ud \omega < \infty.
\end{aligned}
\label{awt4}
\end{equation}

Now, using the analytic wavelet, we consider real and imaginary parts separately as follows: let $\psi(t)=\psi_\text{r}(t)+j\psi_\text{i}(t)$ be an analytic wavelet where imaginary part is the HT or FQT of real part, $\Psi(\omega)=2U(\omega)\Psi_\text{r}(\omega)$, $U(\omega)$ is unit step function, $\psi_{s,\tau}(t)=\psi_{\text{r}\,s,\tau}(t)+j\psi_{\text{i}\,s,\tau}(t)$, then from \eqref{awt1} we can write $W_\psi(s,\tau)=W_{\text{r}\,\psi}(s,\tau)-jW_{\text{i}\,\psi}(s,\tau)$ where $W_{\text{r}\,\psi}(s,\tau)=\langle x(t),\psi_{\text{r}\,s,\tau}(t)\rangle$ and $W_{\text{i}\,\psi}(s,\tau)=\langle x(t),\psi_{\text{i}\,s,\tau}(t)\rangle$, thus we can write
\begin{equation}
\begin{aligned}
x(t)&=\frac{2}{C_\psi} \int_{-\infty}^{\infty} \int_{-\infty}^{\infty} W_{\text{r}\,\psi}(s,\tau) \psi_{\text{r}\,s,\tau}(t) \ud \tau \frac{1}{s^2}\ud s,\\
x(t)&=\frac{2}{C_\psi} \int_{-\infty}^{\infty} \int_{-\infty}^{\infty} W_{\text{i}\,\psi}(s,\tau) \psi_{\text{i}\,s,\tau}(t) \ud \tau \frac{1}{s^2}\ud s,
\end{aligned}
 \label{awt7}
\end{equation}
and define the two representations of wavelet quadrature transform (WQT) of signal $x(t)$ as
\begin{equation}
\begin{aligned}
x_{w}(t)&=\frac{2}{C_\psi} \int_{-\infty}^{\infty} \int_{-\infty}^{\infty} W_{\text{r}\,\psi}(s,\tau) \psi_{\text{i}\,s,\tau}(t) \ud \tau \frac{1}{s^2}\ud s,\\
x_{w}(t)&=\frac{2}{C_\psi} \int_{-\infty}^{\infty} \int_{-\infty}^{\infty} W_{\text{i}\,\psi}(s,\tau) \psi_{\text{r}\,s,\tau}(t) \ud \tau \frac{1}{s^2}\ud s,
\end{aligned}
\label{awt8}
\end{equation}
where $x_{w}(t)=x(t,\pi/2)$ and the subscript $w$ indicates that the phase-shift has been obtained using the wavelet transform. Therefore for the analytic wavelet, using \eqref{awt7} and \eqref{awt8}, original signal $x(t)$ \eqref{awt31} can be written as
\begin{equation}
\begin{aligned}
x(t)&=\frac{1}{C_\psi} \int_{-\infty}^{\infty} \int_{-\infty}^{\infty}\Big([W_{\text{r}\,\psi}(s,\tau) \psi_{\text{r}\,s,\tau}(t)+ W_{\text{i}\,\psi}(s,\tau) \psi_{\text{i}\,s,\tau}(t)] \\
& +j [W_{\text{r}\,\psi}(s,\tau) \psi_{\text{i}\,s,\tau}(t)-W_{\text{i}\,\psi}(s,\tau) \psi_{\text{r}\,s,\tau}(t)] \Big) \ud \tau \frac{1}{s^2}\ud s,\\
&=\frac{1}{C_\psi} \int_{-\infty}^{\infty} \int_{-\infty}^{\infty} [W_{\text{r}\,\psi}(s,\tau) \psi_{\text{r}\,s,\tau}(t)+ W_{\text{i}\,\psi}(s,\tau) \psi_{\text{i}\,s,\tau}(t)].
\end{aligned}
\label{awt81}
\end{equation}

Thus, using these two quadrature component representations \eqref{awt8}, one can define the two representations of wavelet analytic signal (WAS) as
\begin{equation}
\begin{aligned}
z_{w}(t)&=x(t)+jx_{w}(t)=\frac{2}{C_\psi} \int_{-\infty}^{\infty} \int_{-\infty}^{\infty} W_{\text{r}\,\psi}(s,\tau) \psi_{s,\tau}(t) \ud \tau \frac{1}{s^2}\ud s,\\
z_{w}(t)&=x(t)+jx_{w}(t)=\frac{2}{C_\psi} \int_{-\infty}^{\infty} \int_{-\infty}^{\infty} W^*_{\psi}(s,\tau) \psi_{\text{r}\,s,\tau}(t) \ud \tau \frac{1}{s^2}\ud s.
\end{aligned}
\label{awt9}
\end{equation}
%where, in general, $x_{w_1}(t)=x_{w_2}(t)=x(t,\pi/2)$.

Now, we introduce the desired phase in the AWT \eqref{awt4} as
\begin{equation}
W_\psi(s,\tau,\alpha(\omega))=\frac{\sqrt{|s|}}{2\pi}\int_{0}^{\infty} \Psi^{*}(s\omega) X(\omega)e^{j\omega \tau} e^{-j\alpha(\omega)} \ud \omega. \label{awt5}
\end{equation}
If the introduced phase in the AWT \eqref{awt5} is independent of frequency, i.e. $\alpha(\omega)=\alpha$, then we can write $W_\psi(s,\tau,\alpha)=W_\psi(s,\tau)e^{-j\alpha}=[\cos(\alpha)W_{\text{r}\,\psi}(s,\tau)-\sin(\alpha)W_{\text{i}\,\psi}(s,\tau)]-j[\cos(\alpha)W_{\text{i}\,\psi}(s,\tau)+\sin(\alpha)W_{\text{r}\,\psi}(s,\tau)]$, and we obtain two representations of an arbitrary constant WPT of a signal as
\begin{equation}
\begin{aligned}
x(t,\alpha)&=\frac{2}{C_\psi} \int_{-\infty}^{\infty} \int_{-\infty}^{\infty} W_{\text{r}\,\psi}(s,\tau,\alpha) \psi_{\text{r}\,s,\tau}(t) \ud \tau \frac{1}{s^2}\ud s,\\
x(t,\alpha)&=\frac{2}{C_\psi} \int_{-\infty}^{\infty} \int_{-\infty}^{\infty} W_{\text{i}\,\psi}(s,\tau,\alpha) \psi_{\text{i}\,s,\tau}(t) \ud \tau \frac{1}{s^2}\ud s.
\end{aligned}
\label{awt10}
\end{equation}
 
We can also obtain constant WPT of WAS \eqref{awt9} as
\begin{equation}
\begin{aligned}
z_{w}(t,\alpha)&=[x(t)+jx_{w}(t)]e^{-j\alpha}=x(t,\alpha)+jx_{w}(t,\alpha),
%z_{w}(t,\alpha)&=[x(t)+jx_{w}(t)]e^{-j\alpha}=x(t,\alpha)+jx_{w}(t,\alpha+\pi/2),
\end{aligned}
\label{awt11}
\end{equation}
where $x_{w}(t,\alpha)=\cos(\alpha)x_w(t)-\sin(\alpha)x(t)$, and define the WPT of a signal $x(t)$ as
\begin{equation}
x(t,\alpha)=\cos(\alpha)x(t)+\sin(\alpha)x_{w}(t). \label{awt110}
\end{equation}       

\subsubsection{A single-integral representation of the WPT}
A single-integral can be used for the inverse continuous wavelet transform (CWT), if a signal $x(t)$ and a wavelet $\psi(t)$ are satisfying the following two conditions \cite{Matlab1}: (1) the signal $x(t)$ is a real-valued function, and (2) either the wavelet $\psi(t)$ is an even function which has a real-valued Fourier transform, or the wavelet $\psi(t)$ is an analytic wavelet which has Fourier transform $\Psi(\omega)=0$ for $\omega<0$ and supports only for positive frequencies.

If $x(t)$ and $y(t)$ are two finite energy signals, $\psi_1(t)$ and $\psi_2(t)$ are two wavelet functions which satisfy the two-wavelet admissibility condition, the following equality holds:
\begin{equation}
C_{\psi_1,\psi_2}\langle x(t),y(t)\rangle=\int_{-\infty}^{\infty} \int_{-\infty}^{\infty} W_{\psi_1}(s,\tau) W^*_{\psi_2}(s,\tau) \ud \tau \frac{1}{s}\ud s, \label{awt12}
\end{equation}
where $W_{\psi_1}(s,\tau)=\langle x(t),\psi_1(s,\tau)(t)\rangle$, $W_{\psi_2}(s,\tau)=\langle y(t),\psi_2(s,\tau)(t)\rangle$ and
\begin{equation}
C_{\psi_1,\psi_2}=\int_{-\infty}^{\infty} \frac{\Psi^*_1(\omega)\Psi_2(\omega)}{|\omega|} \ud \omega, \label{awt13}
\end{equation}
and the two-wavelet admissibility condition is defined as
\begin{equation}
\int_{-\infty}^{\infty} \frac{|\Psi^*_1(\omega)||\Psi_2(\omega)|}{|\omega|} \ud \omega<\infty. \label{awt13_1}
\end{equation}

The main idea of single-integral inverse CWT is that admissibility condition of the two-wavelet can be satisfied even if either one of the wavelets is not admissible, and it can be further simplified by allowing one of the signals and wavelets to be distributions. Thus, by considering analytic wavelet $\psi_1(t)$, real-valued signal $x(t)$, both signal $y(t)$ and wavelet $\psi_2(t)$ to be the delta function, one can obtain the WAS representation using the following single-integral inverse CWT as
\begin{equation}
z(t)=x(t)+jx_w(t)=\frac{2}{C_{\psi_1,\delta}}\int_{0}^{\infty} W_{\psi_1}(s,t) \frac{1}{s}\ud s, \label{awt14}
\end{equation}
where imaginary part $x_w(t)$ is the WQT of signal $x(t)$, and if $W_{\psi_1}(s,\tau,\alpha)=W_{\psi_1}(s,\tau)e^{-j\alpha}$ as defined in \eqref{awt5}, then we define
\begin{equation}
z(t,\alpha)=x(t,\alpha)+jx_w(t,\alpha)=\frac{2}{C_{\psi_1,\delta}}\int_{0}^{\infty} W_{\psi_1}(s,t,\alpha) \frac{1}{s}\ud s, \label{awt15}
\end{equation}
where $x(t,\alpha)$ is the proposed WPT which introduces desired phase-shift $\alpha$ in the signal $x(t)$ using AWT, $x(t,0)=x(t)$, $x(t,\pi/2)=x_w(t)$ and $x_w(t,\alpha)=x(t,\alpha+\pi/2)$.   
%%%%%%%%%%%%%%%%%%%%%%%%%%%%%%%%%

\subsection{Implementation of the GFR using DFT}\label{DFT_GFR}
The DFT and inverse DFT (IDFT) of a signal $x[n]$ of length $N$ are defined as
\begin{equation}
\begin{aligned}
X[k]&=\frac{1}{N} \sum_{n=0}^{N-1} x[n]\exp(-j {2\pi kn}/{N}), \quad 0 \le k \le N-1, \\
x[n]&=\sum_{k=0}^{N-1} X[k]\exp(j {2\pi kn}/{N}), \quad 0 \le k \le N-1.
\end{aligned}\label{DFT1}
\end{equation}
The DFT and IDFT are computed efficiently using the fast Fourier transform (FFT) algorithm. Unless otherwise mentioned, we consider $x[n]$ as a real valued signal. We obtain the discrete-time PT kernel corresponding to continuous-time kernel \eqref{FS1} as
\begin{equation}
\hbar[n,\alpha]= \cos(\alpha)\delta[n]+\sin(\alpha)h[n], \label{FSd}
\end{equation} 
where $\hbar[n,\alpha]=\delta[n,\alpha]$, $\delta[n,0]=\delta[n]$, and $h[n]=\hat{\delta}[n]=\delta[n,\pi/2]=\frac{(1-\cos(\pi n))}{\pi n}$ is the discrete-time HT kernel.
Thus we can obtain a constant phase shift in a signal $x[n]$ as $x[n,\alpha]= \cos(\alpha)x[n]+\sin(\alpha)\hat{x}[n]$, where $x[n]=x[n,0]$ and HT $\hat{x}[n]=x[n,\pi/2]$.
The PT defined in \eqref{eq4} and \eqref{FT4} can be computed using FFT by considering real part of analytic signal
\begin{equation}
z[n,\alpha_k]=x[n,\alpha_k]+jx[n,\alpha_k+\pi/2]=\text{IFFT}\{X[k]H[k]\}, \label{fftcomp1}
\end{equation} 
i.e., $x[n,\alpha_k]=\text{Re}\{z[n,\alpha_k]\}$, where $H[k]$ is defined (if $N$ is even) as 
\begin{equation}
H[k] =
\begin{cases}
\cos(\alpha_k),  \quad k=0,N/2,\\
2 \exp{(-j\alpha_k)},  \quad 1 \le k \le N/2-1,\\
%\exp{(-j\alpha_k)}, \quad k=N/2, \\
0, \quad (N/2+1) \le k \le N-1,
\end{cases}\label{zna}
\end{equation}
or (if $N$ is odd)
\begin{equation}
H[k] =
\begin{cases}
\cos(\alpha_k),  \quad k=0, \\
2 \exp{(-j\alpha_k)},  \quad 1 \le k \le (N-1)/2,\\
0, \quad ((N-1)/2+1) \le k \le N-1.
\end{cases}\label{zna1}
\end{equation}
For a constant phase shift, imaginary part is the HT of real part in \eqref{fftcomp1}, and if $\alpha_k=0, \forall k$, real part is the original signal $x[n]$.
In \eqref{zna} and \eqref{zna1}, we can remove the multiplication factor of 2 and define $H[k]=\exp{(-j\alpha_k)}$ for positive frequencies (e.g., $1 \le k \le N/2-1$) and its complex conjugate $H[k]=\exp{(j\alpha_k)}$ for negative frequencies (e.g., $(N/2+1) \le k \le N-1$).   

\textbf{Observation 3.3.1:}\label{obj2} The phase shift of a sinusoidal signal does not change its amplitude (or energy), except for lowest (i.e. DC) and highest frequency components. For example, if $x[n]=c+\cos(\pi n)$, then $x[n,\alpha]=\cos(\alpha)(c+\cos(\pi n))$, and for $\alpha=\pi/2$, it becomes zero which cannot net be recovered by further phase shift. So, to overcome this issue, we can define phase shift of a constant signal and highest frequency component as $z[n,\alpha]=(c+\cos(\pi n))\,e^{-j\alpha}=c\,e^{-j\alpha}+e^{j(n\pi-\alpha)}$, which preserves the energy in these cases as well, and in the case of $\pi/2$ phase shift, complete signal is getting transfered to imaginary part, and we can recover original signal by further phase shift of ($\alpha+2m\pi$), and $x[n,\alpha]=\text{Re}\{z[n,\alpha]\}$. This is also consistent with complex plane representation, where multiplication of $\pm j$ with a complex number ($z=a+jb$) introduces $\pm \pi/2$ phase shift (i.e. $jz=ja-b$) without any change in amplitude. Therefore, we can use $H[k]=e^{-j\alpha}$ for $k=0,N/2$ in \eqref{zna} and \eqref{zna1}.        

We observe that a simply delayed signal $x[n-n_0]$ obtained using IDFT \eqref{DFT1} as 
\begin{equation}
x[n-n_0]=X[0]+\sum_{k=1}^{N-1} X[k]\exp(j {2\pi k(n-n_0)}/{N}), \label{delay1}
\end{equation}
is valid only for an integer value of $n_0$, because complex conjugate of $\exp(\frac{-j2\pi k n_0}{N})$ is $\exp(\frac{-j2\pi (N-k)n_0}{N})$ only for some integer value of $n_0$, and it is not valid for fractional value of delay. 
Therefore, in order to obtain both integer and fractional delay $n_k\in \mathbb{R}$ in a signal $x[n]$, corresponding to \eqref{eq6} which can be computed by \eqref{fftcomp1}, we define $H[k]$ as
\begin{equation}
H[k] =
\begin{cases}
1,  \quad k=0,\\
2 \exp{(-j2\pi k n_k/N)},  \quad 1 \le k \le N/2-1,\\
\exp{(-j\pi n_k)},  \quad k= N/2,\\
0, \quad (N/2+1) \le k \le N-1,
\end{cases}\label{delay2}
\end{equation}
where $N$ is even (similarly it can also we defined when $N$ is odd) and for some constant delay $n_0$ one can set $n_k=n_0, \forall k$.  

The $\mu$-th order derivative approximation of a signal $x[n]$, denoted as $x_\mu[n]$, corresponding to \eqref{eq7}, can be estimated by
\begin{equation}
x_\mu[n]=a_0\frac{(n/F_s)^{-\mu}}{\Gamma \left( 1-\mu \right)}+\text{Re}\big[\text{IFFT}\{X[k]H[k]\}\big] \label{fdfft}
 \end{equation}
using $H[k]$ which we define as
\begin{equation}
H[k] =
\begin{cases}
0,  \quad k=0,\\
2 (2\pi k/N)^{\mu} \exp{(j\mu \pi /2)},  \quad 1 \le k \le N/2-1,\\
(\pi)^{\mu} \exp{(j \mu \pi/2)},  \quad k= N/2,\\
0, \quad (N/2+1) \le k \le N-1,
\end{cases}\label{diffft}
\end{equation}
where mean-value $a_0=\sum_{n=0}^{N-1}x[n]/N$, $\mu\ge 0$, and $N$ is even (similarly it can also we defined when $N$ is odd); and when $\mu\le 0$ then it is $\mu$-th order integral of a signal as defined in \eqref{eq8}.

\subsection{Implementation of GFR using DCT}\label{DCT_GFR}
 
The DCT-2 of a sequence, $x[n]$ of length $N$, is defines as \cite{IEEECompt} 
\begin{equation}
X_{\text{c}2}[k]=\sqrt{\frac{2}{N}} \sigma_k \sum_{n=0}^{N-1} x[n] \cos\left(\frac{\pi k (2n+1)}{2N}\right), \qquad 0\le k \le N-1, \label{fdct}
\end{equation}
and inverse DCT (IDCT) is obtained by
\begin{equation}
x[n]= \sqrt{\frac{2}{N}} \sum_{k=0}^{N-1} \sigma_k {X_{\text{c}2}[k]} \cos\left(\frac{\pi k (2n+1)}{2N}\right), \qquad 0\le n \le N-1, \label{idct}
\end{equation}
where normalization factors $\sigma_k=\frac{1}{\sqrt{2}}$ for $k=0$, and $\sigma_k=1$ for $k\ne 0$. If consecutive samples of a sequence $x[n]$ are correlated, then DCT concentrates energy in a few $X_{\text{c}2}[k]$ and decorrelates them. The DCT basis sequences, $\cos\left(\frac{\pi k (2n+1)}{2N}\right)$, which are a class of discrete Chebyshev polynomials \cite{IEEECompt}, form an orthogonal set as inner product $\left<\cos\left(\frac{\pi k (2n+1)}{2N}\right), \cos\left(\frac{\pi m (2n+1)}{2N}\right)\right>=0$ for $k\ne m$.

The discrete Fourier cosine quadrature transform (FCQT), $\tilde{x}_{\text{c}2}[n]$, of a signal $x[n]$ is defined as \cite{FQT}
 \begin{equation}
 \tilde{x}_{\text{c}2}[n]= \sqrt{\frac{2}{N}}\sum_{k=0}^{N-1} {X_{\text{c}2}[k]} \sin\left(\frac{\pi k (2n+1)}{2N}\right), \qquad 0\le n \le N-1, \label{fqt}
 \end{equation}
where $X_{\text{c}2}[k]$ is the DCT-2 of a signal $x[n]$.
We define the PT using the DCT-2 as 
\begin{equation}
x[n,\alpha(k)]= \sqrt{\frac{2}{N}} \sum_{k=0}^{N-1} \sigma_k {X_{\text{c}2}[k]} \cos\left(\frac{\pi k (2n+1)}{2N}-\alpha(k)\right), \qquad 0\le n \le N-1, \label{dct_fspt}
\end{equation}
and if $\alpha(k)=\alpha$, then we can write \eqref{dct_fspt} as
\begin{equation}
x[n,\alpha]= \cos(\alpha)x[n]+\sin(\alpha)\tilde{x}_{\text{c}2}[n]. \label{dct_fspt1}
\end{equation}
Moreover, a  variable fractional time-delay in a signal can be introduced using \eqref{dct_fspt} as $x[n-n_k]=x[n,\alpha(k)]$, with $\alpha(k)={\pi k n_k}/{N}$ where $n_k\in \mathbb{R}$, and for a constant fractional time-delay we set $n_k=n_0, \forall k$.

Using 8-types of DCTs and 8-types of DSTs, 16-types of Fourier quadrature transforms (FQTs) are defined in \cite{FQT}. Therefore, we can obtained 16-types of PT using the DCTs/DSTs and 16-types of FQTs. Here, we have presented the PT, in \eqref{dct_fspt} and \eqref{dct_fspt1}, using only DCT-2 and corresponding FQT, rest other 15-types of PTs can be easily defined in a very similar way.

\section{Results and discussions}\label{simre}
In this section, we present simulation results to demonstrate the efficacy of the proposed methods GFR and PT. We mainly consider those signals which have been widely used in literature for performance evaluation and results comparison among the proposed and other existing methods.  

\begin{figure}[!t]
	\centering
	\begin{tabular}{c}
		%\sidesubfloat[]{\includegraphics[angle=0,width=0.47\textwidth,height=0.4\textwidth]{Gaussian_phaseShift_0_pi}}	
		%\sidesubfloat[]{\includegraphics[angle=0,width=0.47\textwidth,height=0.4\textwidth]{Gaussian_phaseShift_pi_2pi}}
		{\includegraphics[angle=0,width=0.5\textwidth,height=0.35\textwidth]{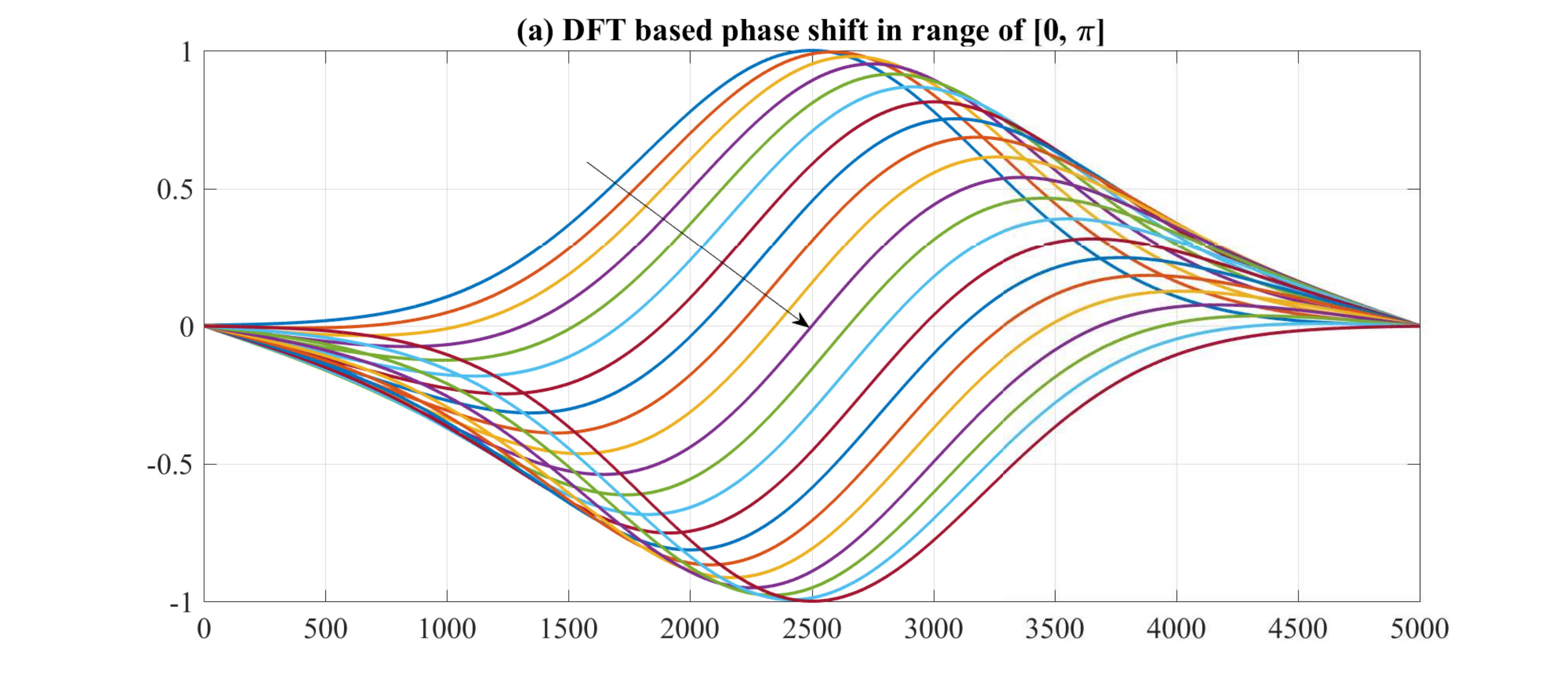}}
		{\includegraphics[angle=0,width=0.5\textwidth,height=0.35\textwidth]{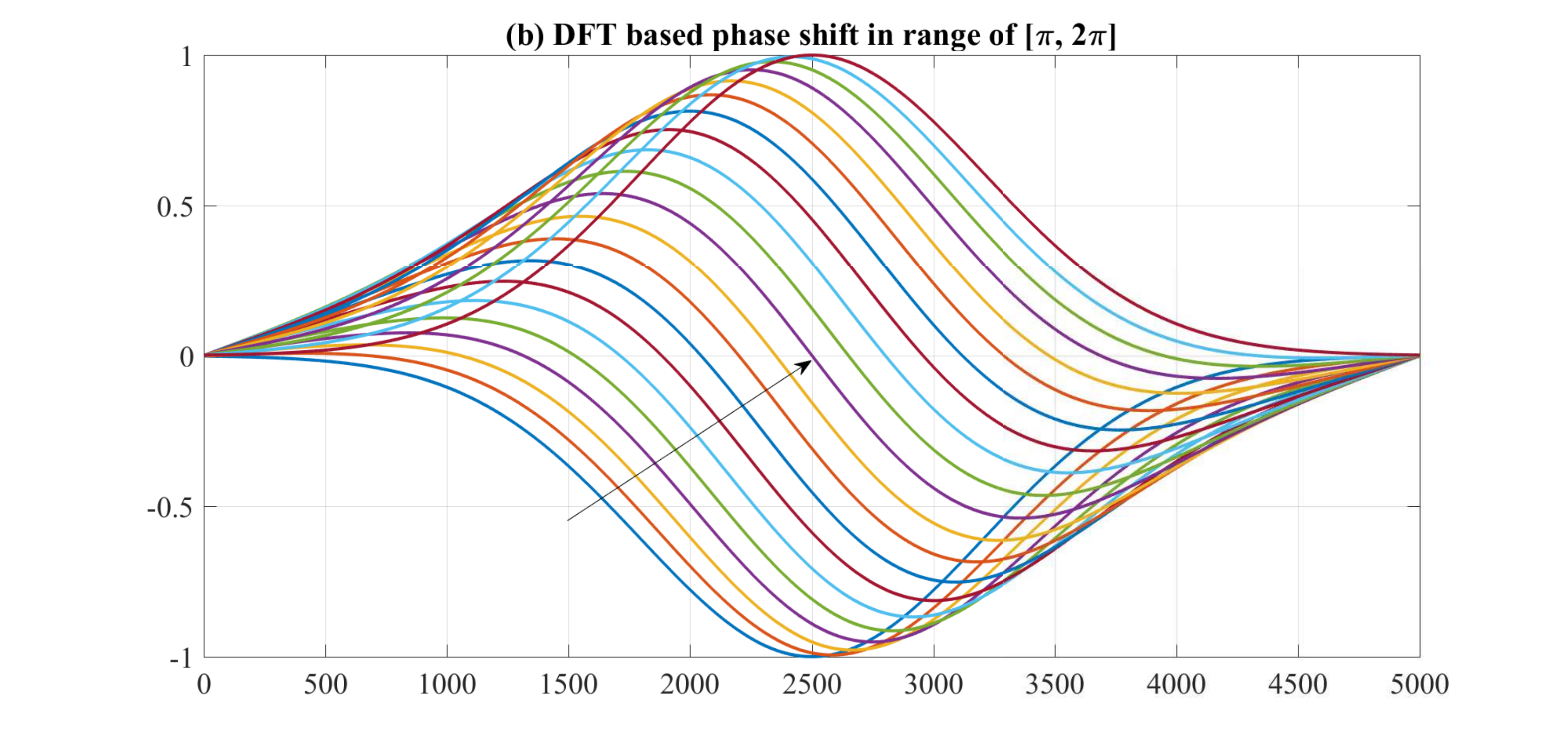}} \\
		{\includegraphics[angle=0,width=0.5\textwidth,height=0.35\textwidth]{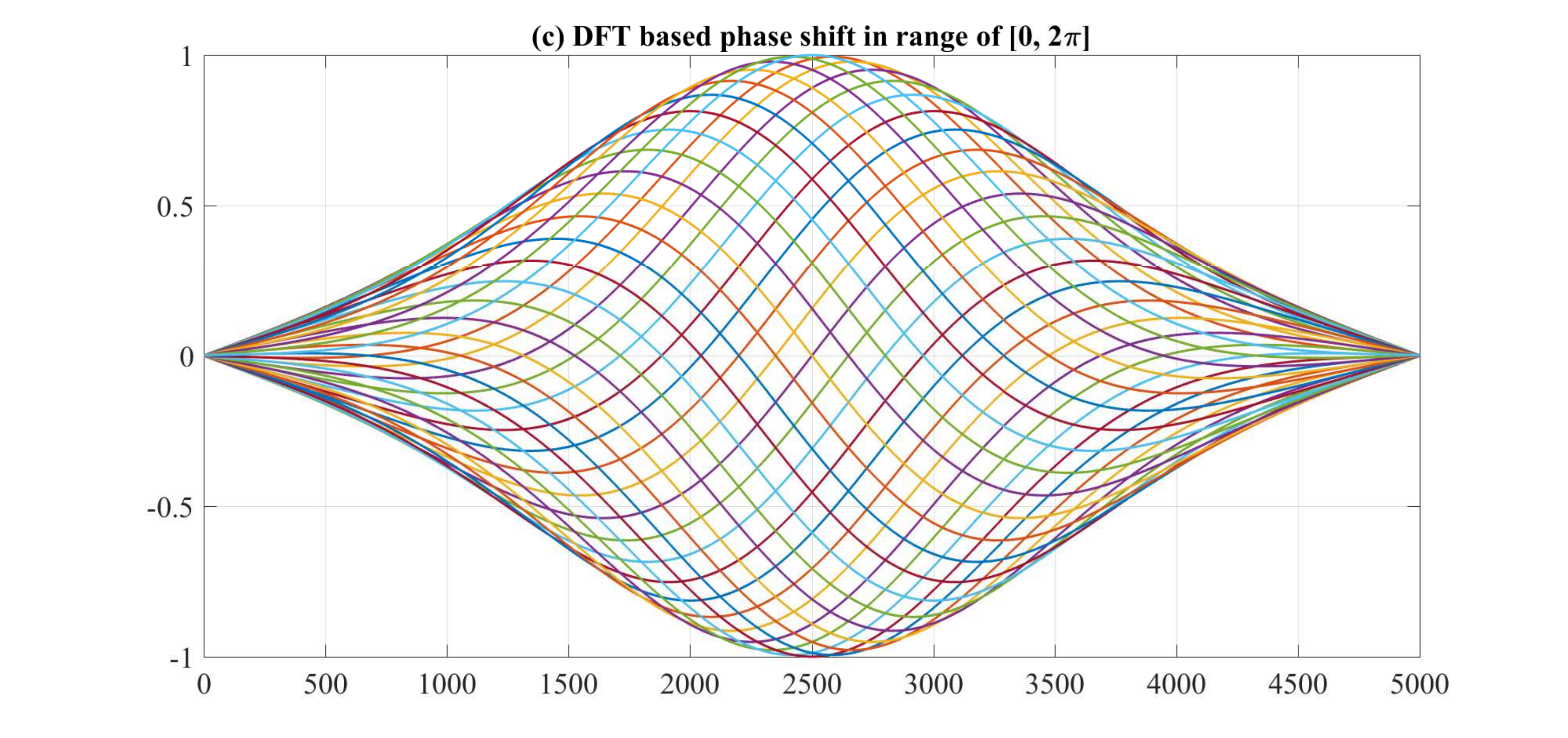}}
		{\includegraphics[angle=0,width=0.5\textwidth,height=0.35\textwidth]{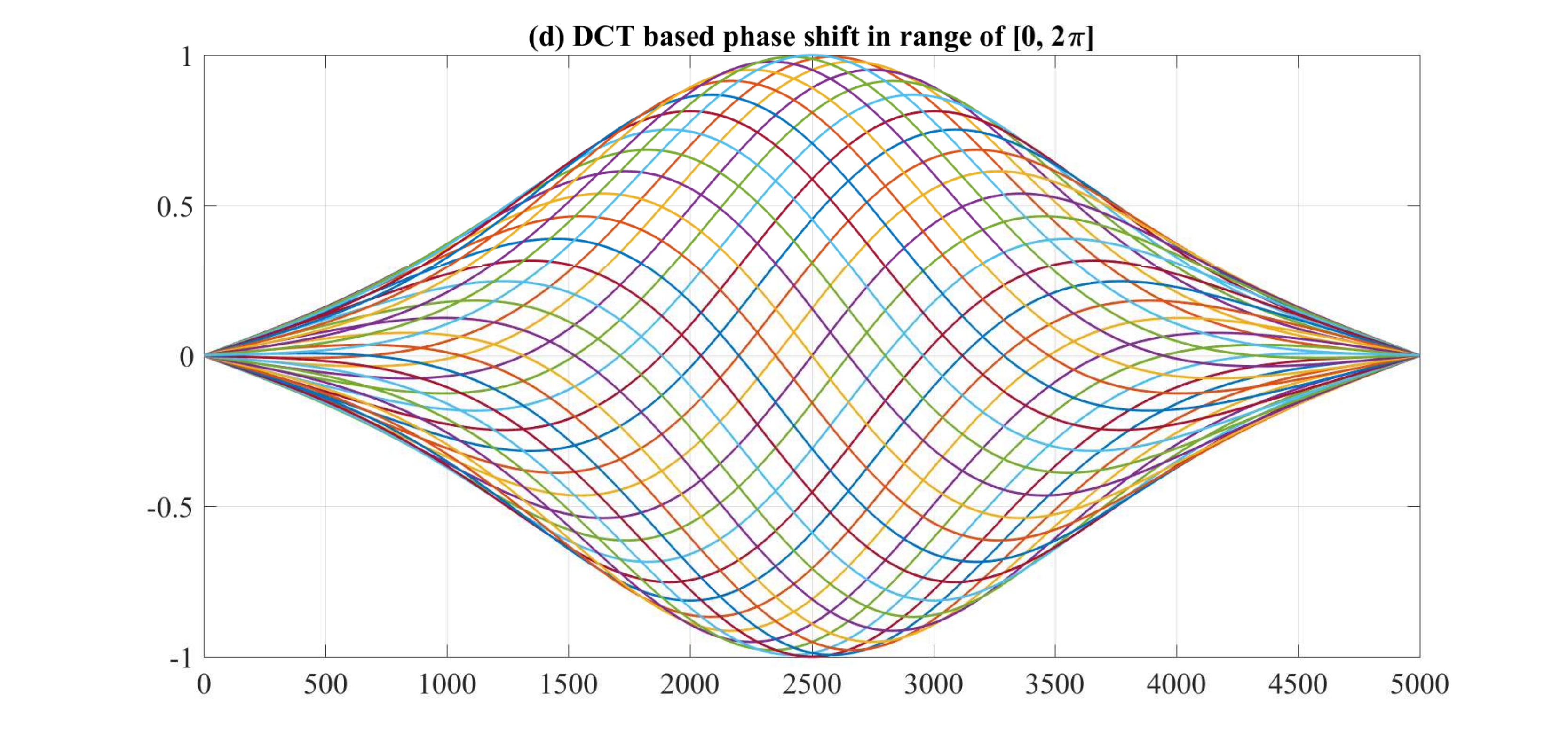}}	
	\end{tabular}
	\caption{Phase shift analysis of a Gaussian function of Example 1: (a) Phase in the range of $[0, \pi]$ is increasing in a step of $\pi/20$, in the direction of arrow, first plot is original Gaussian function and last one corresponds to $\pi$ phase-shift, plot corresponding to the tip of arrow is the HT of original signal, i.e. $\pi/2$ phase shift, (b) Phase in the range of $[\pi, 2\pi]$ is increasing in a step of $\pi/20$, in the direction of arrow, first plot corresponds to $\pi$ phase-shift and last one is original Gaussian function obtained by $2\pi$ phase shift, plot corresponding to the tip of arrow is the HT of original signal with minus sign, i.e. $3\pi/2$ phase shift. Phase shift in the range of $[0, 2\pi]$ using DFT (c), and using DCT (d).}\label{fig:FSAS_GAS_Gaussian}
\end{figure}

\textbf{Example 1:}
In this example, we consider phase shift analysis of a Gaussian function, $x(t)=e^{-(t-2.5)^2}$, $0\le t<5$ with sampling frequency $F_s=1000$ Hz, which is shown in Figure~\ref{fig:FSAS_GAS_Gaussian}, where in the direction of arrow (a) Phase in the range of $[0, \pi]$ radians is increasing in step of $\pi/20$ radian, first plot is original Gaussian function and last one corresponds to $\pi$ radian phase-shift, plot corresponding to the tip of arrow is the HT of original signal, i.e. $\pi/2$ radian phase shift, (b) Phase in the range of $[\pi, 2\pi]$ radians is increasing in step of $\pi/20$ radian, first plot corresponds to $\pi$ radian phase-shift and last one is original Gaussian function obtained with $2\pi$ phase shift, plot corresponding to the tip of arrow is the HT of original signal with minus sign, i.e. $3\pi/2$ radian phase shift, (c) Phase shift in the range of $[0, 2\pi]$ using DFT which is obtained by combining (a) and (b); (d) Phase shift in the range of $[0, 2\pi]$ using DCT. We observe that there is no difference, in phase shift obtained by the DFT \eqref{FS2} and DCT \eqref{fqtc11} approaches, for a set of signals which represent same underlying periodic extension that inherently present in DFT ($N$-sample periodicity) and DCT ($2N$-sample periodicity with even symmetry) representations.

\begin{figure}[!t]
	\centering
	\begin{tabular}{c}
		%\sidesubfloat[]{\includegraphics[angle=0,width=0.47\textwidth,height=0.4\textwidth]{Sin_phaseShift_0_pi_GAS}}	
		%\sidesubfloat[]{\includegraphics[angle=0,width=0.47\textwidth,height=0.4\textwidth]{Sin_phaseShift_pi_2pi_GAS}}
		{\includegraphics[angle=0,width=0.5\textwidth,height=0.35\textwidth]{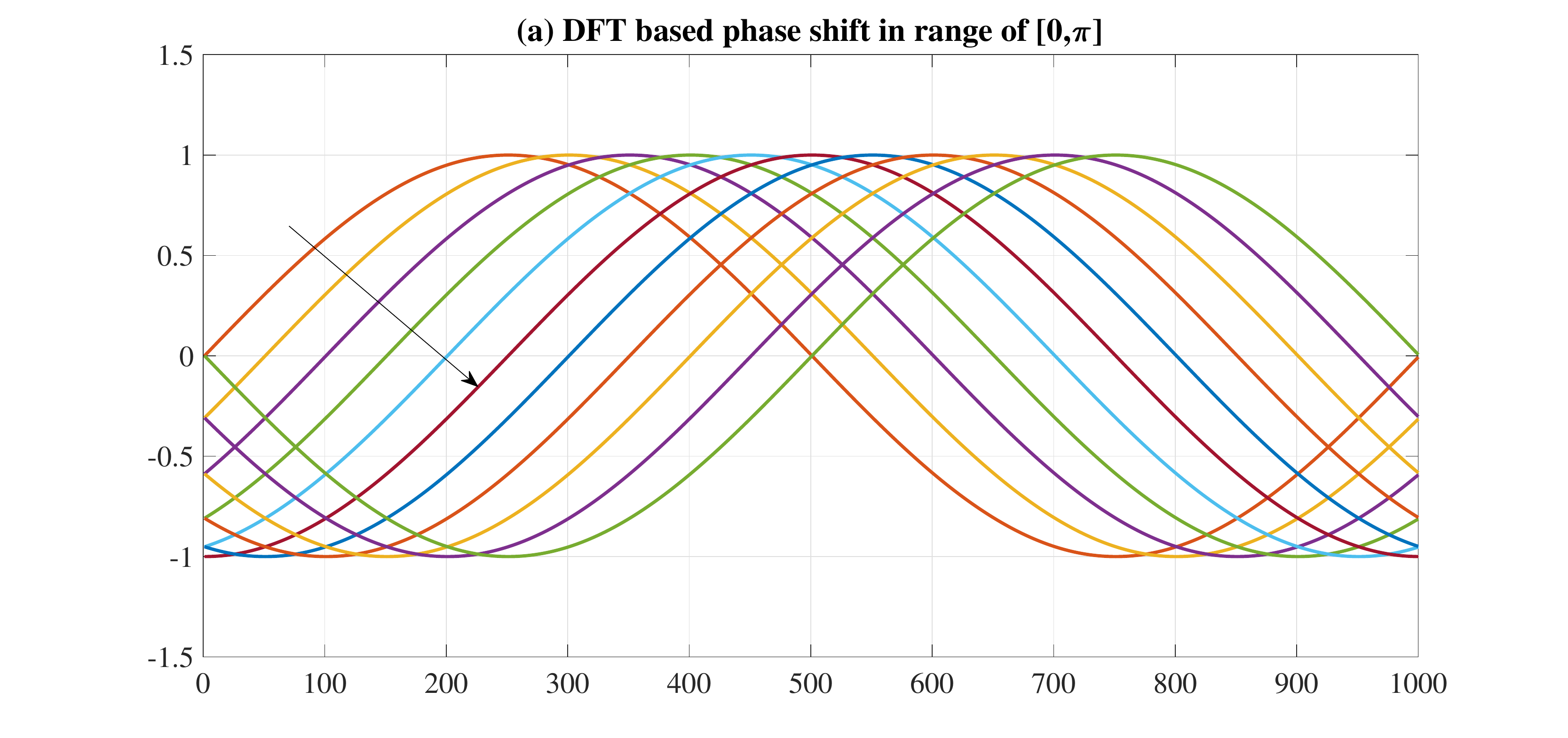}}
		{\includegraphics[angle=0,width=0.5\textwidth,height=0.35\textwidth]{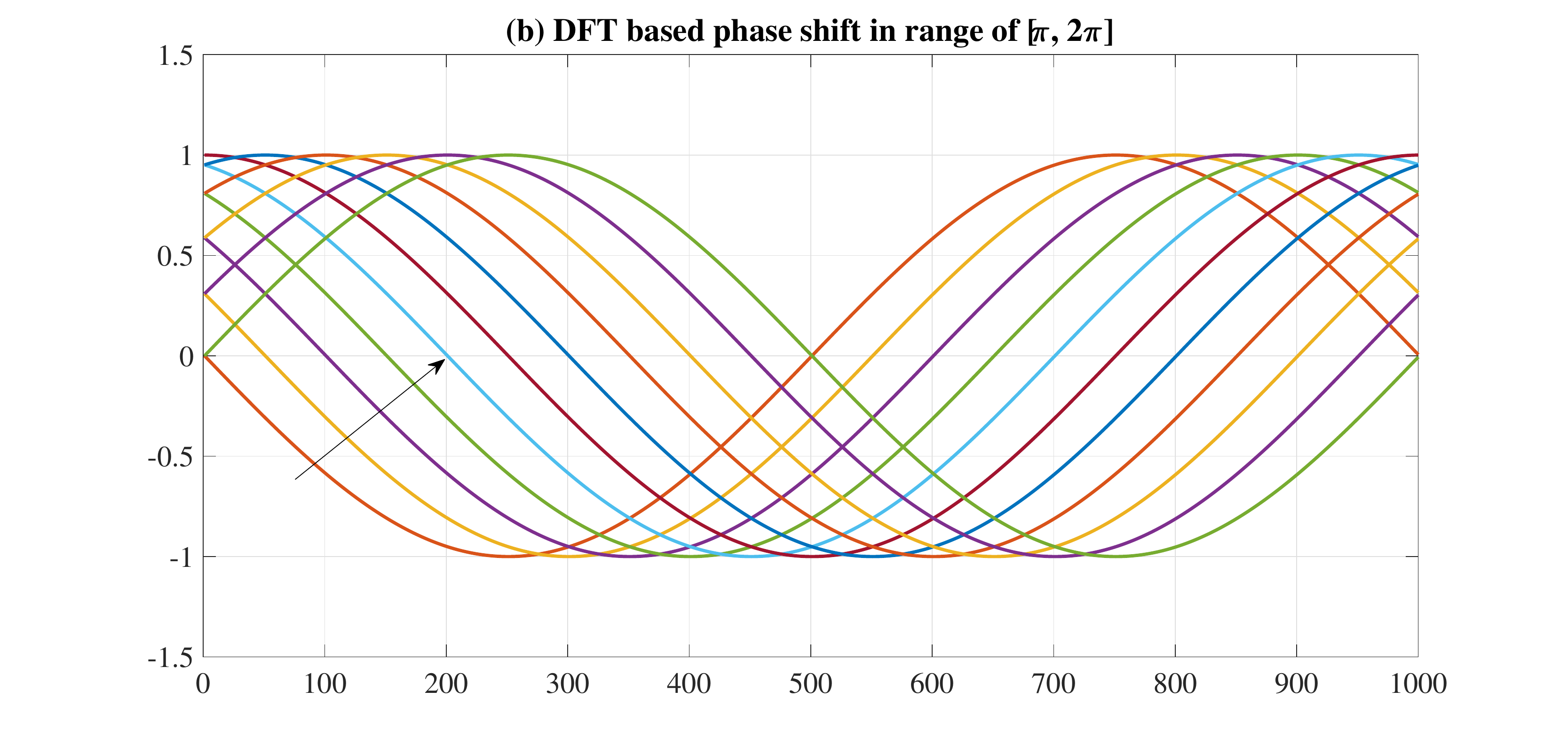}} \\	
		{\includegraphics[angle=0,width=0.5\textwidth,height=0.35\textwidth]{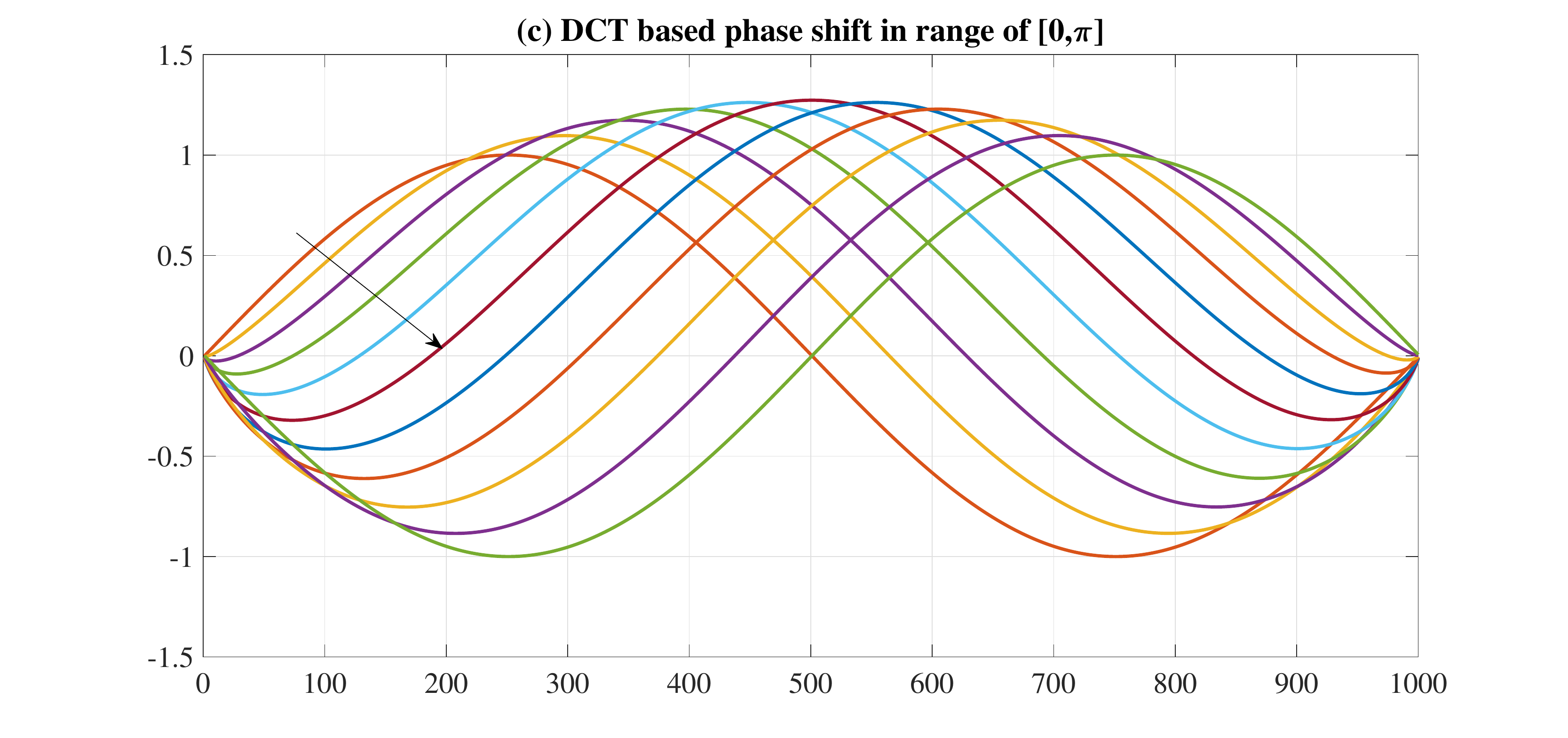}}
		{\includegraphics[angle=0,width=0.5\textwidth,height=0.35\textwidth]{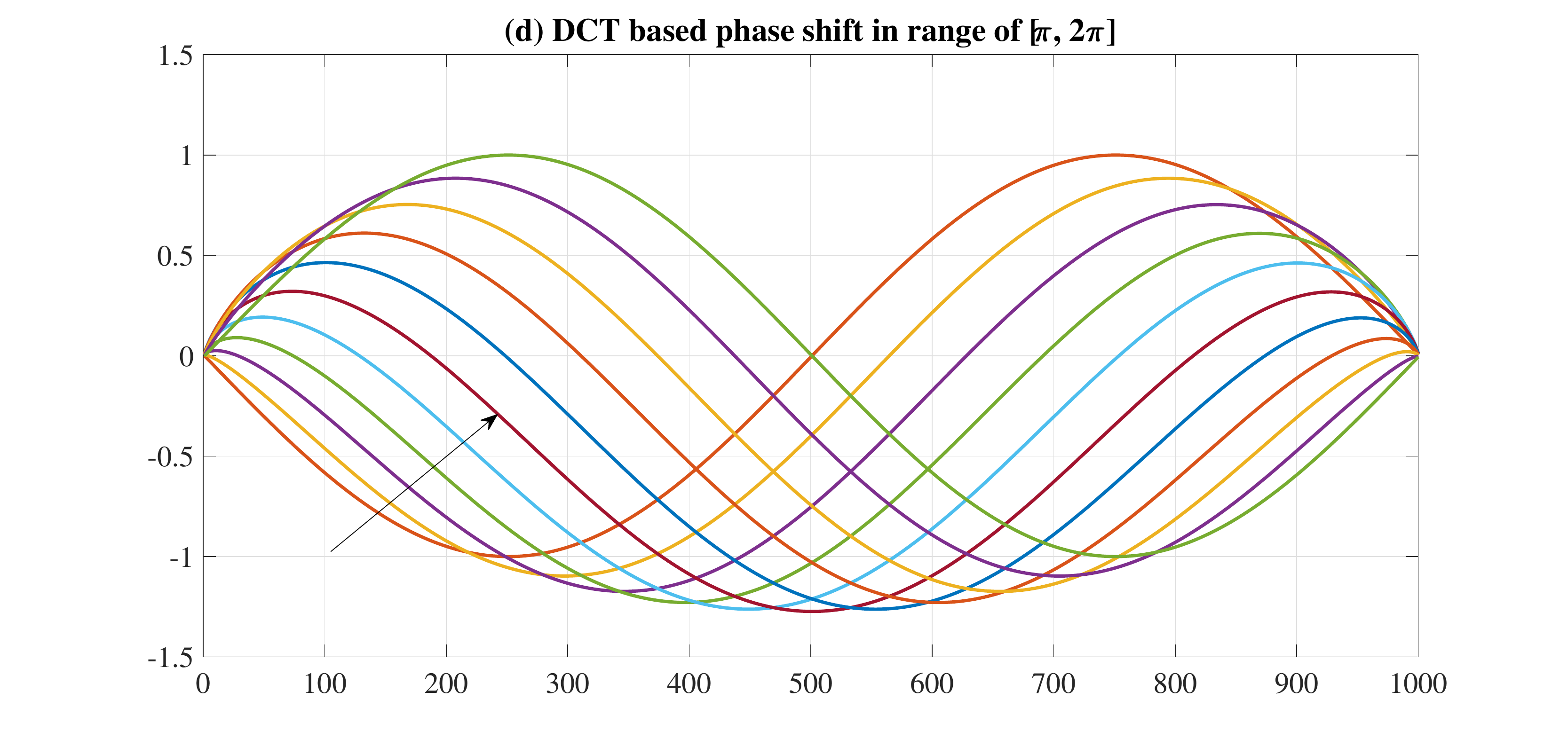}} \\
		{\includegraphics[angle=0,width=0.5\textwidth,height=0.35\textwidth]{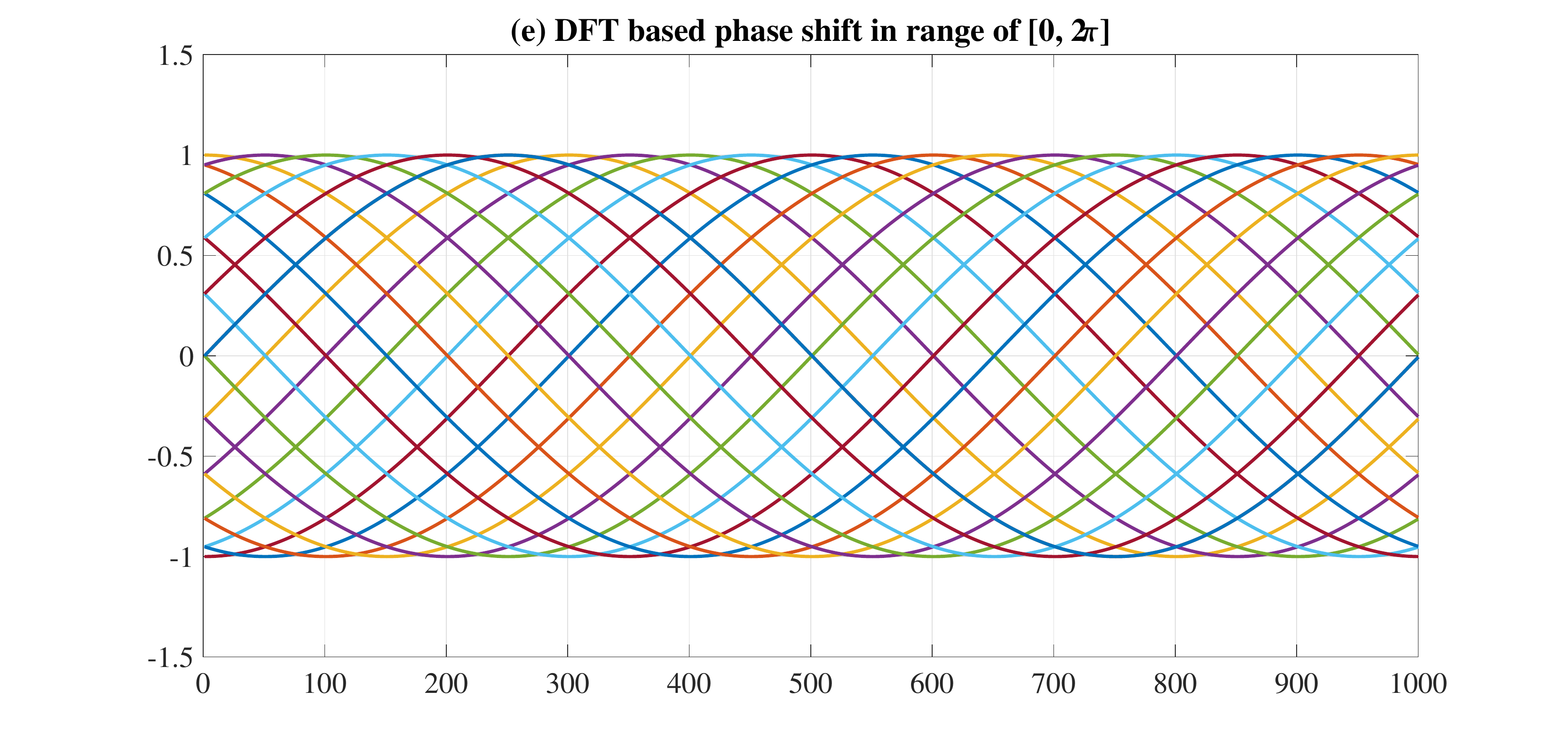}}
		{\includegraphics[angle=0,width=0.5\textwidth,height=0.35\textwidth]{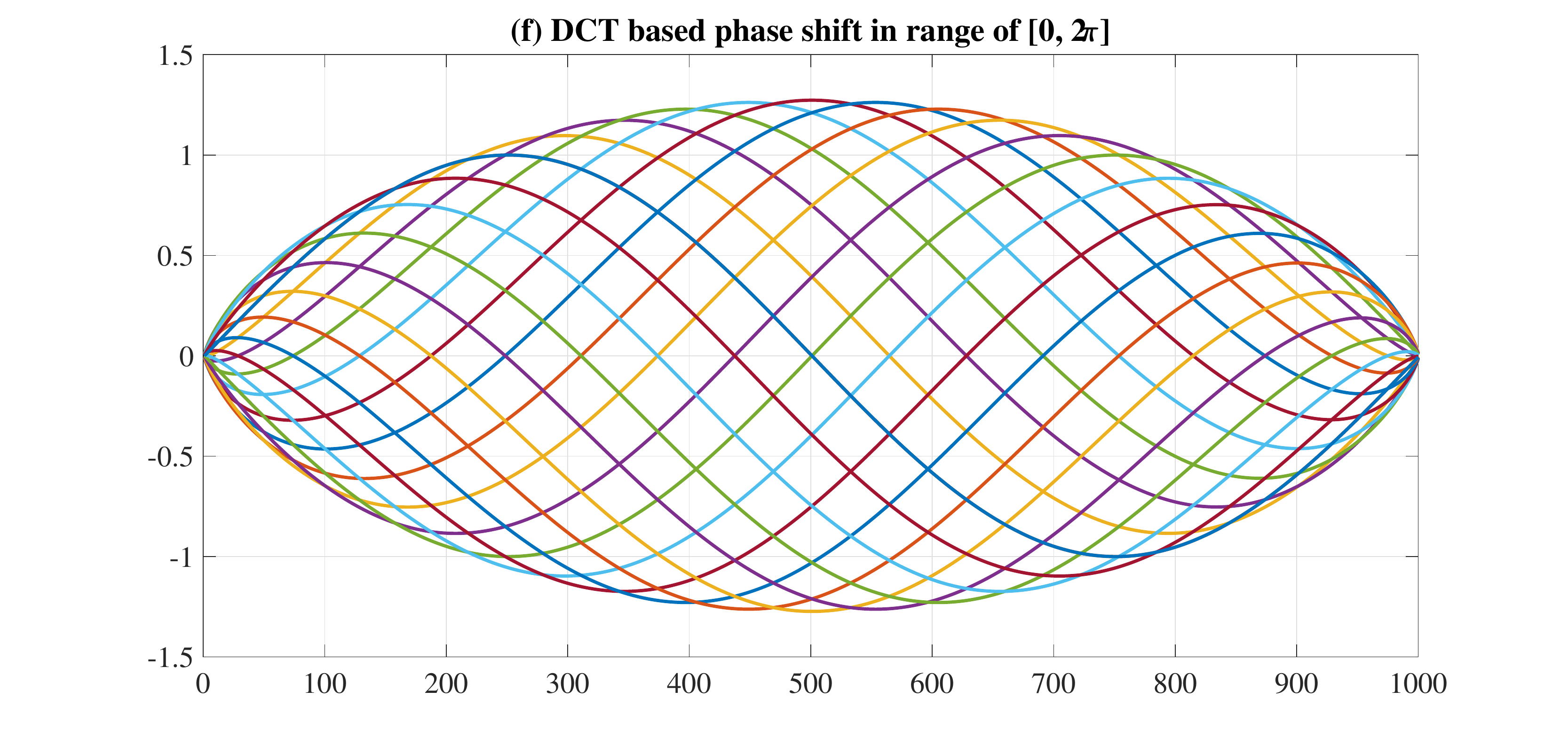}}			
	\end{tabular}
	\caption{Phase shift analysis of the sine function of Example 2 where phase is increasing in step of $\pi/10$: using DFT (a) and DCT (c) with phase in the range of $[0, \pi]$, in the direction of arrow, first plot is original sine function and last one corresponds to $\pi$ phase-shift, plot corresponding to the tip of arrow is $\pi/2$ phase shift; using DFT (b) and DCT (d) with phase in the range of $[\pi, 2\pi]$, in the direction of arrow, first plot is $\pi$ phase shifted sine function and last one corresponds to $2\pi$ phase-shift, plot corresponding to the tip of arrow is $3\pi/2$ phase shift; using DFT (e) and DCT (f) with phase in the range of $[\pi, 2\pi]$.}\label{fig:FSAS_GAS_SIN}
\end{figure}

\textbf{Example 2:} Figure \ref{fig:FSAS_GAS_SIN} presents phase shift analysis of sine function, $x(t)=\sin(2\pi t)$, with $0 \le t < 1$ and $F_s=1000$ Hz, where phase is increasing in step of $\pi/10$ radians: (a) using DFT and (c) using DCT with phase in the range of $[0, \pi]$, in the direction of arrow, first plot is original sine function and last one corresponds to $\pi$ phase-shift, plot corresponding to the tip of arrow is the $\pi/2$ phase shift; (b) using DFT and (d) using DCT with phase in the range of $[\pi, 2\pi]$, in the direction of arrow, first plot is $\pi$ phase shifted sine wave and last one corresponds to $2\pi$ phase-shift, plot corresponding to the tip of arrow is $3\pi/2$ phase-shift; (e) using DFT which is obtained by superimposing (a) and (b); and (f) using DCT which is obtained by superimposing (c) and (d). We observe clear differences, in phase shift obtained by the DFT \eqref{FS2} and DCT \eqref{fqtc11} approaches, for a set of signals which yield one periodic signal for DFT ($N$-sample periodicity) representation and another periodic signal for DCT ($2N$-sample periodicity with even symmetry) representation.  

\begin{figure}[!t]
	\centering	
		\includegraphics[angle=0,width=1\textwidth,height=0.5\textwidth]{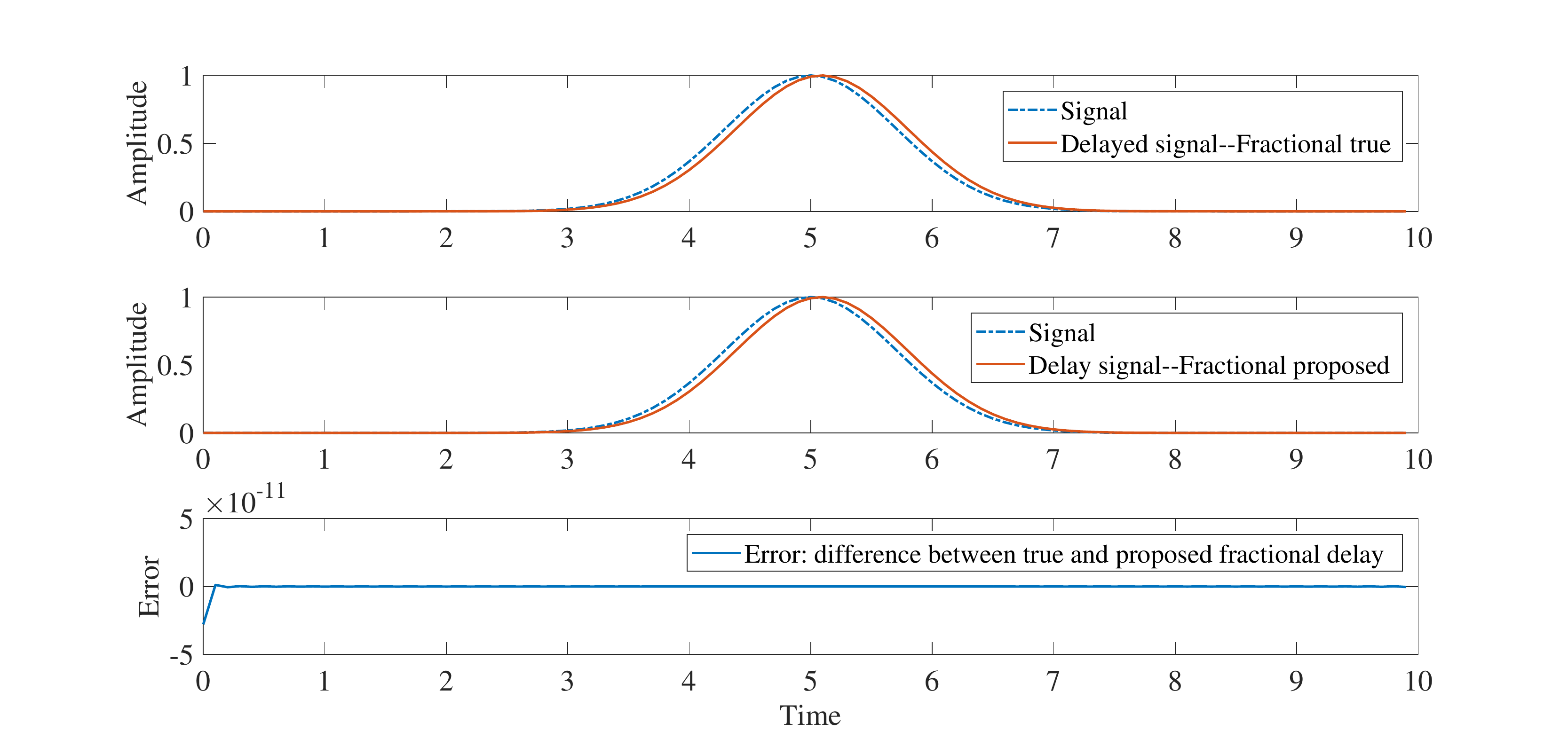}	
	\caption{Fractional delay analysis of a Gaussian function of Example 3: (upper) Original Gaussian function and its delayed version obtained theoretically, (middle) Original Gaussian function and its delayed version obtained by proposed method, and (lower) Error estimated by taking the difference between truly delayed signal and delayed signal obtained by proposed method.}\label{fig:FDGaussian}
\end{figure}

\textbf{Example 3:} In this example, we consider a Gaussian function $x(t)=e^{-(t-5)^2}$, with $0 \le t < 10$, $F_s=1/T=10$ Hz, and thus $x[n]=e^{-(nT-5)^2}$. We computed true delayed signal as $x(t-t_0)=e^{-(t-t_0-5)^2}$, and delayed signal $x[n-n_0]$ using the proposed method \eqref{delay2} with fractional delay $t_0=n_0 T$ sec, where $n_0=0.9$. Figure \ref{fig:FDGaussian} shows Fractional delay estimation of this Gaussian function: (upper) Original Gaussian function and its delayed version obtained theoretically using expression $x(t-t_0)=e^{-(t-t_0-5)^2}$, (middle) Original Gaussian function and its delayed version obtained by proposed method \eqref{delay2}, and (lower) Estimated error, which is really small in order of $10^{-11}$, by taking the difference between truly delayed signal and delayed signal obtained by proposed method.

\begin{figure}[!t]
	\centering	
	\includegraphics[angle=0,width=1\textwidth,height=0.5\textwidth]{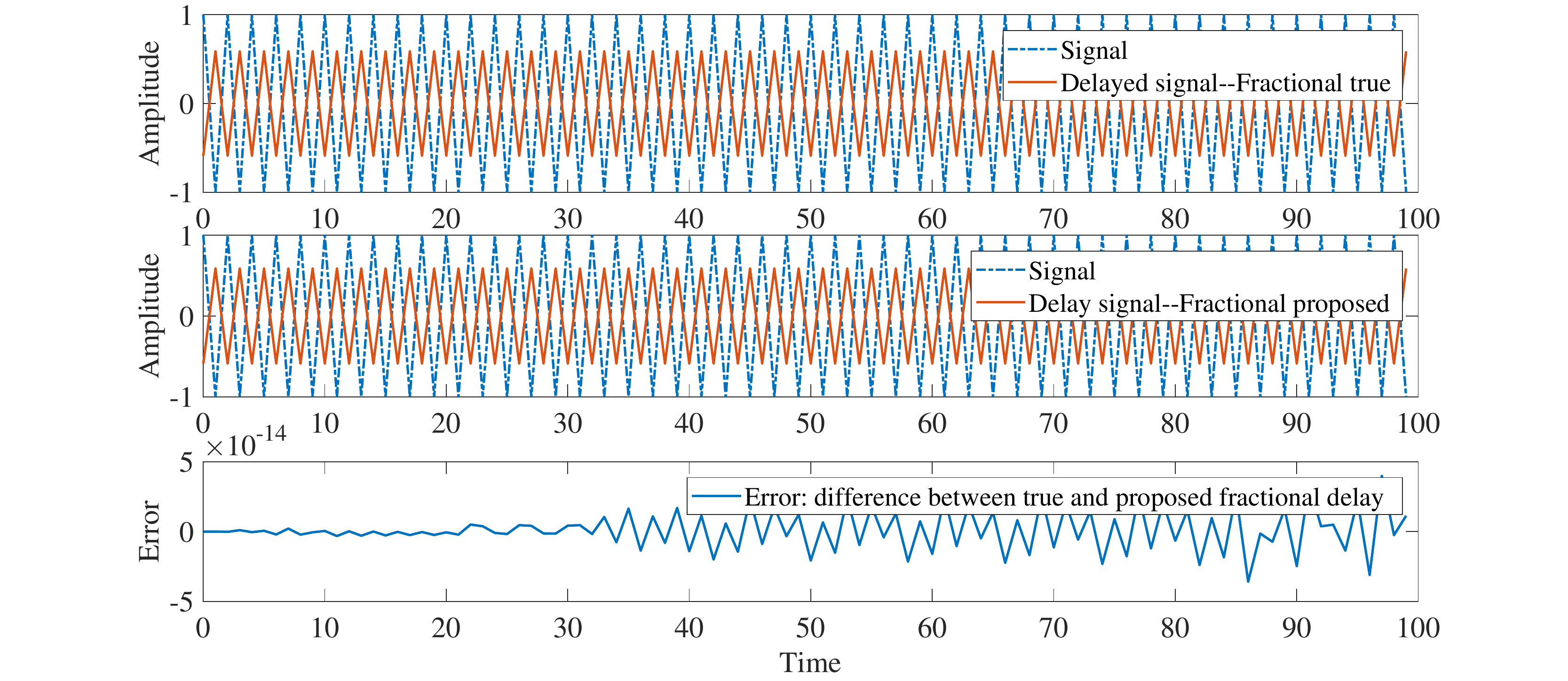}	
	\caption{Fractional delay analysis of $\cos(\pi t)$ function of Example 3: (upper) Original cos function and its delayed version obtained theoretically, (middle) Original cos function and its delayed version obtained by proposed method, and (lower) Error estimated by taking the difference between truly delayed signal and delayed signal obtained by proposed method.}\label{fig:FDCos}
\end{figure}

We consider a cos function $x(t)=\cos(\pi t)$, with $0 \le t < 100$, $F_s=1/T=1$ Hz, and thus $x[n]=\cos(\pi n)$. We computed true delayed signal as $x(t-t_0)=\cos(\pi (t-t_0))$, and delayed signal $x[n-n_0]$ using the proposed method \eqref{delay2} with fractional delay $t_0=n_0 T$ sec, where $n_0=0.7$. Figure \ref{fig:FDCos} shows Fractional delay estimation of this cos function: (upper) Original cos function and its delayed version obtained theoretically using expression $x(t-t_0)=\cos(\pi (t-t_0))$, (middle) Original cos function and its delayed version obtained by proposed method \eqref{delay2}, and (lower) Estimated error, which is really small in order of $10^{-14}$, by taking the difference between truly delayed signal and delayed signal obtained by proposed method.

%\begin{figure}[!t]
%	\centering	
%	\includegraphics[angle=0,width=1\textwidth,height=0.5\textwidth]{cos_pit_FD}	
%	\caption{Fractional delay analysis of a cos function of Example 3: (upper) Original cos function and its delayed version obtained theoretically, (middle) Original cos function and its delayed version obtained by proposed method, and (lower) Error estimated by taking the difference between truly delayed signal and delayed signal obtained by proposed method.}\label{fig:FDCos1}
%\end{figure}

\begin{figure}[!t]
	\centering	
	\includegraphics[angle=0,width=1\textwidth,height=0.5\textwidth]{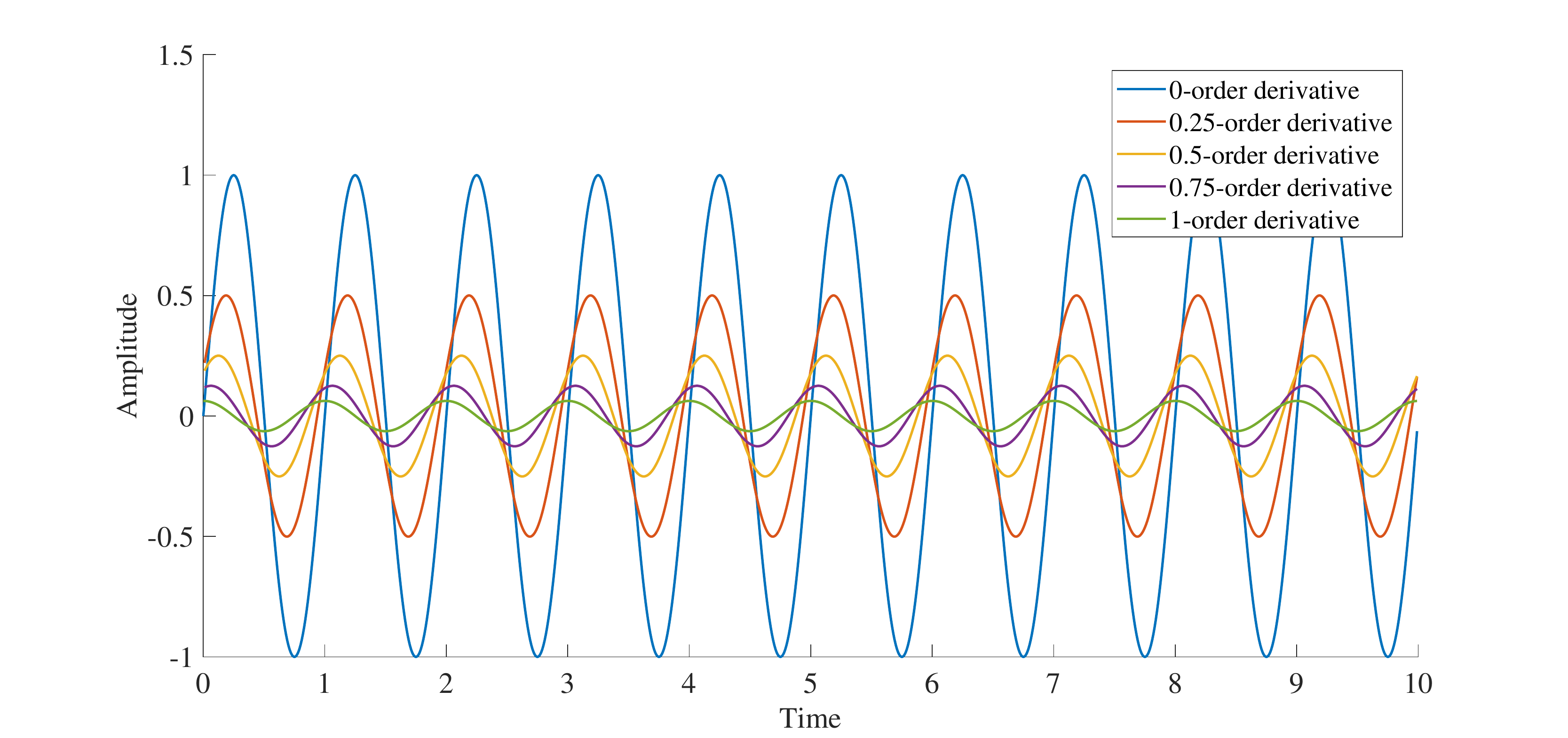}	
	\caption{Estimation of fractional derivative of order $0.0,0.25,0.5,0.75$ and $1.0$ of a sine function of Example 4 by proposed method.}\label{fig:FDerSine}
\end{figure}

\begin{figure}[!t]
	\centering	
	\includegraphics[angle=0,width=1\textwidth,height=0.5\textwidth]{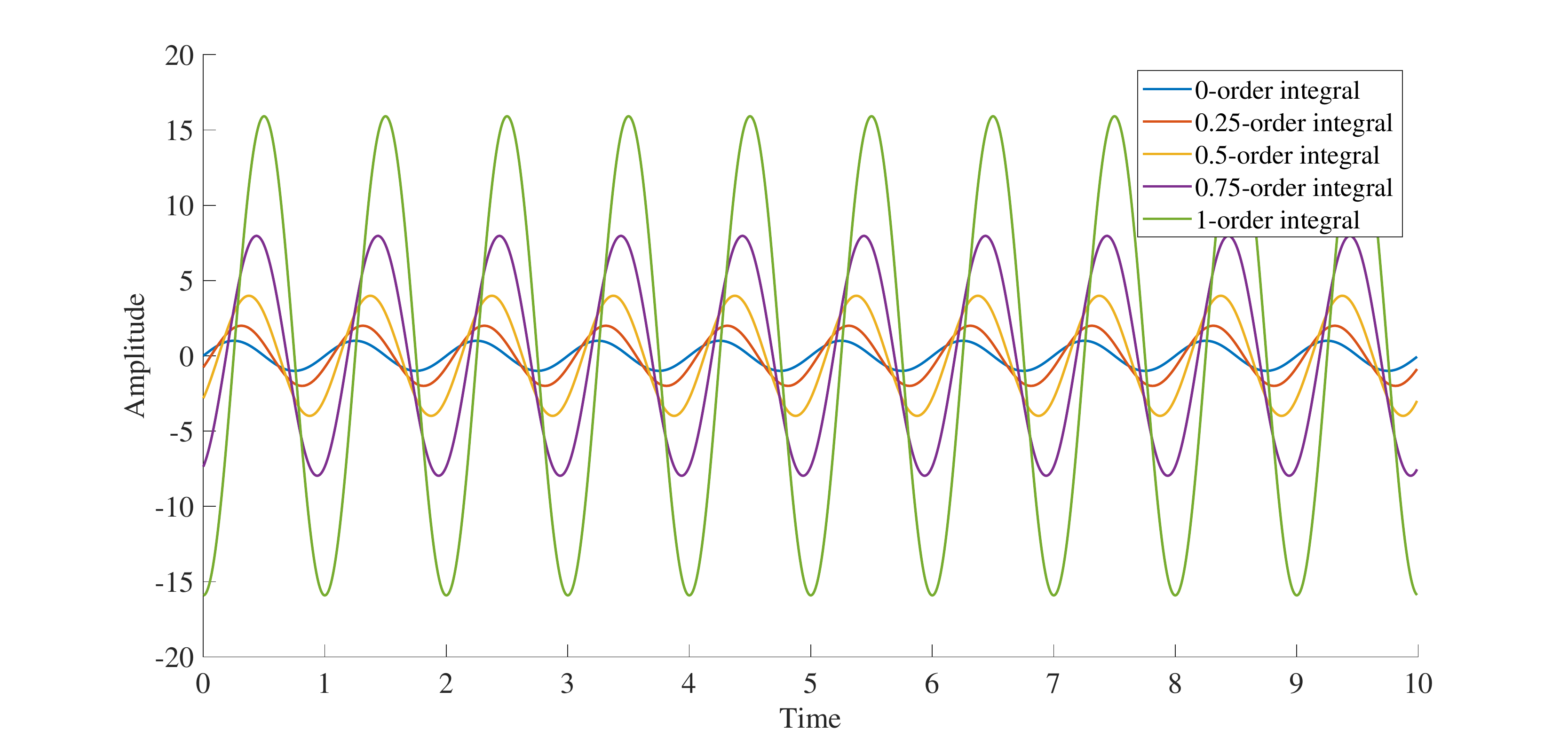} 	
	\caption{Estimation of fractional integral of order $0.0,0.25,0.5,0.75$ and $1.0$ of a sine function of Example 4 by proposed method.}\label{fig:FIntSine}
\end{figure}

\textbf{Example 4:}  In this example, we consider fractional order derivative (FOD) and fractional integral (FOI) of sine function, $x(t)=\sin(2\pi t)$, with $0 \le t < 10$ and $F_s=1000$ Hz. Figure \ref{fig:FDerSine} and Figure \ref{fig:FIntSine} present the FOD and FOI, respectively, of sine wave where fractional order $\mu\in \{0.0,0.25,0.5,0.75, 1.0\}$; FD and FI are estimated using proposed method \eqref{fdfft}. 


\begin{figure}[!t]
	\centering
	\includegraphics[angle=0,width=1\textwidth,height=0.5\textwidth]{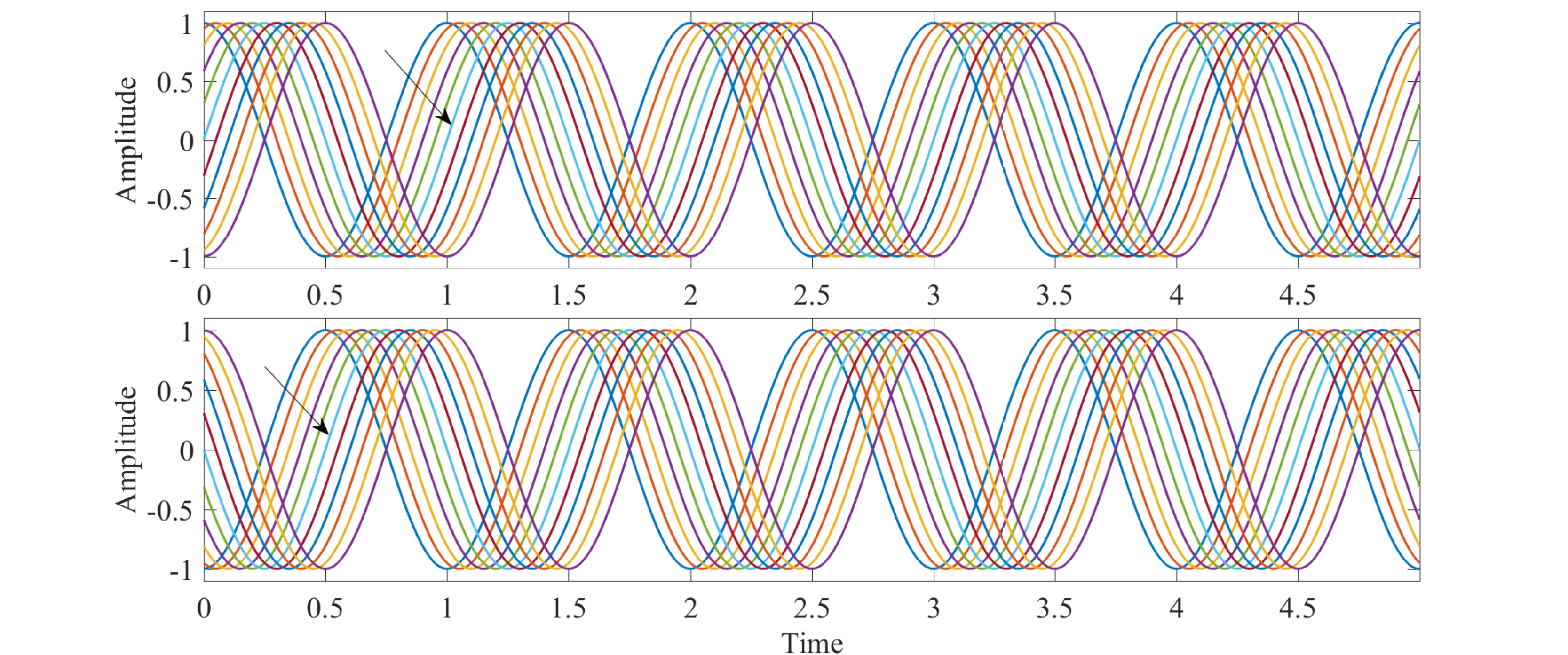}
	\caption{Phase shift analysis of the cosine function of Example 5 where phase is increasing in step of $\pi/10$: using the proposed WPT with Morse wavelet (a) top plot has phase in the range of $[0, \pi]$, in the direction of arrow, first plot is original cosine function and last one corresponds to $\pi$ phase-shift, plot corresponding to the tip of arrow is $\pi/2$ phase shift; (b) bottom plot has phase in the range of $[\pi, 2\pi]$, in the direction of arrow, first plot is $\pi$ phase shifted cosine function and last one corresponds to $2\pi$ phase-shift, plot corresponding to the tip of arrow is $3\pi/2$ phase shift.}\label{fig:WPT_SIN}
\end{figure}

\textbf{Example 5:} In this example, we consider a cosine wave and obtain desired phase shift using the proposed wavelet phase transform (WPT) and wavelet quadrature transform (WQT). Figure \ref{fig:WPT_SIN} presents phase shift analysis of cosine function, $x(t)=\cos(2\pi t)$, with $0 \le t < 5$ and $F_s=1000$ Hz, where phase is increasing in step of $\pi/10$ radians using the proposed WPT with Morse wavelet: (a) top plot with phase in the range of $[0, \pi]$, in the direction of arrow, first plot is original cosine function and last one corresponds to $\pi$ phase-shift, plot corresponding to the tip of arrow is the $\pi/2$ phase shift (i.e. WQT); (b) bottom plot has phase in the range of $[\pi, 2\pi]$, in the direction of arrow, first plot is $\pi$ phase shifted cosine wave and last one corresponds to $2\pi$ phase-shift, plot corresponding to the tip of arrow is $3\pi/2$ phase-shift in the original signal.  

\section{Conclusion}\label{con}
In this work, we introduced the generalized Fourier representation (GFR) which is completely based on the Fourier representation of a signal, and studied seven special cases of the GFR, namely Fourier representation, phase transform (PT), time-delay including fractional delay of discrete time signals, time derivative and integral including fractional order, amplitude modulation and frequency modulation.
The most important and fundamental contribution of this study is the PT which is a special case of the GFR and a true generalization of the Hilbert transform. Using the proposed PT, the desired phase-shift and time-delay can be obtained in a signal under analysis. We derived PT kernel to obtain any constant phase shift, discussed the various properties of the PT, and showed that the HT is a special case of PT when phase-shift is $\pi/2$ radian. We also provided an extension of the one-denominational PT for two-dimensional image signals in Appendix \ref{MDPSPT}, which can easily be extended for higher dimensional signals. Using the PT, we demonstrated that (i) a constant phase shift (e.g. $\pi/2$ phase shift) in a signal corresponds to variable time-delays in all harmonics, (ii) a frequency dependent phase shift in all harmonics of Fourier representation can be used to obtain a constant time-delay in a signal, (iii) a constant phase shift is same as the constant time-delay only for single frequency sinusoid. 
The time derivative and time integral, including fractional order, of a signal are obtained using the GFR. We proposed to use DCT based implementation to avoid end artifacts due to discontinuities present in both end of the signal. We proposed to obtained a fractional delay in a discrete time signal using the Fourier representations, i.e. DFT, DSTs and DCTs. We also presented the fast FFT implementation of all the proposed representations. 
Using the analytic wavelet transform (AWT), we proposed wavelet phase transform (WPT) to introduce a desired phase-shift in a signal under-analysis, and presented the two representations of wavelet quadrature transform (WQT) as a special case of the WPT where phase-shift is $\pi/2$ radians.

%\section*{ACKNOWLEDGMENTS}
%Author would like to express his sincere appreciation to the anonymous reviewers for their valuable suggestions. Author also would like to show his gratitude to the Prof. SD Joshi (IITD), Prof. RK Pateny (IITD) and Dr. Kaushik Saha (CTO, Samsung R\&D Institute India--Delhi) for sharing their wisdom and expertise during this research.
\appendix
\section{Multidimensional PT}\label{MDPSPT}
In this appendix, we consider the PT of 2D images which can be easily extended for multidimensional signals. Let $g(x,y)$ be a non-periodic and real function, then the 2D-FT is defined as
\begin{equation}
G(\omega_1,\omega_2)= \int_{-\infty}^{\infty}\int_{-\infty}^{\infty} g(x,y) e^{-j(\omega_1x+\omega_2y)} \ud x \ud y,
\end{equation}
and inverse 2D-FT is defined as
\begin{equation}
g(x,y)= \frac{1}{4\pi^2} \int_{-\infty}^{\infty}\int_{-\infty}^{\infty} G(\omega_1,\omega_2) e^{j(\omega_1x+\omega_2y)} \ud \omega_1 \ud \omega_2.
\end{equation}
The 2D PT transfer function corresponding to 1D \eqref{FSPT_TF} can be written as  
\begin{equation}
H(\alpha(\omega_1,\omega_2)) =
\begin{cases}
e^{-j\alpha(\omega_1, \, \omega_2)},  \quad 0\le \omega_1 < \infty, -\infty <\omega_2< \infty \\
e^{j\alpha(\omega_1, \, \omega_2)},  \quad -\infty< \omega_1 \le 0, -\infty <\omega_2< \infty,
\end{cases}\label{2FSPT_TF}
\end{equation}
where 2D analytic signal (2D-AS) is defined by considering first and fourth quadrants of 2D-FT plane as \cite{rslc11}     
\begin{equation}
z(x,y)= g(x,y)+j\hat{g}(x,y)= \frac{1}{2\pi^2} \int_{-\infty}^{\infty}\int_{0}^{\infty} G(\omega_1,\omega_2) e^{j(\omega_1x+\omega_2y)} \ud \omega_1 \ud \omega_2 \label{2FSPT1}, 
\end{equation}
where $\hat{g}(x,y)$ is the HT of ${g}(x,y)$. The 2D-PT can be computed by considering the real part of the 2D-PT of 2D-AS which we defined as
\begin{equation}
z(x,y,\alpha(\omega_1, \, \omega_2))= \frac{1}{2\pi^2} \int_{-\infty}^{\infty}\int_{0}^{\infty} G(\omega_1,\omega_2) e^{j(\omega_1x+\omega_2y)} e^{-j\alpha(\omega_1, \, \omega_2)} \ud \omega_1 \ud \omega_2, \label{2FSPT2}
\end{equation}
and if $\alpha(\omega_1, \, \omega_2)=\alpha$, then $z(x,y,\alpha)=z(x,y)e^{-j\alpha}$, therefore 2D counter part of 1D PT \eqref{FS2} is defined as
\begin{equation}
g(x,y,\alpha)=\text{Re}\{z(x,y,\alpha)\}= \cos(\alpha)g(x,y)+ \sin(\alpha)\hat{g}(x,y), \label{2FSPT3}
\end{equation}
where $\text{Re}\{z(x,y,\alpha)\}$ denotes real part of the function ${z(x,y,\alpha)}$. As the 2D-AS \eqref{2FSPT1} is not unique and it can also be defined by considering first and second quadrants of 2D-FT plane \cite{rslc11}, so corresponding modifications (i.e. integration limits would be $-\infty <\omega_1< \infty, 0\le \omega_2 < \infty$) can be easily applied to equations \eqref{2FSPT_TF}, \eqref{2FSPT1} and \eqref{2FSPT2}.         

Now, using the Observation \ref{obj1}, to obtain unique 2D-PT and 2D-AS, we define 2D phase shifter in frequency domain as $H(\alpha(\omega_1,\omega_2))=e^{-j\,\text{sign} (\omega_1 + \omega_2)\alpha(\omega_1, \, \omega_2)}$, i.e. 
%\begin{equation}
%H(\alpha(\omega_1,\omega_2)) =
%\begin{cases}
%e^{-j\alpha(\omega_1, \, \omega_2)},  \quad -\omega_2 \le \omega_1 < \infty, -\infty <\omega_2< \infty \\
%e^{j\alpha(\omega_1, \, \omega_2)},  \quad -\infty< \omega_1 \le -\omega_2, -\infty <\omega_2< \infty,
%\end{cases}\label{2FSPT_True}
\begin{equation}
H(\alpha(\omega_1,\omega_2)) =
\begin{cases}
e^{-j\alpha(\omega_1, \, \omega_2)}, \quad \omega_1 + \omega_2\ge 0 \implies -\omega_2 \le \omega_1 < \infty, -\infty <\omega_2< \infty, \\
%\cos(\alpha), \quad \omega_1 + \omega_2=0,\\ 
e^{j\alpha(\omega_1, \, \omega_2)},  \quad \omega_1 + \omega_2<0 \implies -\infty< \omega_1 < -\omega_2, -\infty <\omega_2< \infty,
\end{cases}\label{2FSPT_True}
\end{equation}
where we define 2D analytic signal (2D-AS) by considering of 2D-FT plane with $\omega_1+\omega_2\ge 0$ as     
\begin{equation}
z(x,y)= g(x,y)+j\hat{g}(x,y)= \frac{1}{2\pi^2} \int_{-\infty}^{\infty}\int_{-\omega_2}^{\infty} G(\omega_1,\omega_2) e^{j(\omega_1x+\omega_2y)} \ud \omega_1 \ud \omega_2 \label{2FSPTn1}, 
\end{equation}
where $\hat{g}(x,y)$ is the HT of ${g}(x,y)$. The 2D-PT can be computed by considering the real part of the 2D-PT of 2D-AS which we defined as
\begin{equation}
z(x,y,\alpha(\omega_1, \, \omega_2))= \frac{1}{2\pi^2} \int_{-\infty}^{\infty}\int_{-\omega_2}^{\infty} G(\omega_1,\omega_2) e^{j(\omega_1x+\omega_2y)} e^{-j\alpha(\omega_1, \, \omega_2)} \ud \omega_1 \ud \omega_2, \label{2FSPTn2}
\end{equation}
and if $\alpha(\omega_1, \, \omega_2)=\alpha$, then $z(x,y,\alpha)=z(x,y)e^{-j\alpha}$, therefore 2D counter part of 1D PT \eqref{FS2} is defined as
\begin{equation}
g(x,y,\alpha)=\text{Re}\{z(x,y,\alpha)\}= \cos(\alpha)g(x,y)+ \sin(\alpha)\hat{g}(x,y), \label{2FSPTn3}
\end{equation}
where $\text{Re}\{z(x,y,\alpha)\}$ denotes real part of the function ${z(x,y,\alpha)}$. Thus, we have obtained the 2D-AS \eqref{2FSPTn1} and 2D-PT \eqref{2FSPTn3} which are uniquely defined.
\subsection{Phase Transform of image signal}
The 2D discrete time Fourier transform (2D-DTFT) and inverse 2D-DTFT, for a non-periodic and real-valued signal $g[m,n]$, are defined as
\begin{equation}
G(\Omega_1,\Omega_2)=\sum_{m=-\infty}^{\infty} \sum_{n=-\infty}^{\infty} g[m,n] \exp(-j[\Omega_1 m+\Omega_2 n])  \label{FDM_eq21}
\end{equation}
\begin{equation}
g[m,n]=\frac{1}{2\pi}\frac{1}{2\pi}\int_{-\pi}^{\pi}\int_{-\pi}^{\pi} G(\Omega_1,\Omega_2) \exp(j[\Omega_1 m+\Omega_2 n]) \ud \Omega_1 \ud \Omega_2. \label{FDM_eq22}
\end{equation}
Let $g[m,n]=\delta[m,n]$, then $G(\Omega_1,\Omega_2)=1$ and from \eqref{FDM_eq22}, we can write
\begin{equation}
\delta[m,n]=\frac{1}{2\pi}\frac{1}{2\pi}\int_{-\pi}^{\pi}\int_{-\pi}^{\pi} \exp(j[\Omega_1 m+\Omega_2 n]) \ud \Omega_1 \ud \Omega_2, \label{FDM_eq23}
\end{equation}
which can be further simplified as
\begin{equation}
\delta[m,n]=\frac{1}{2\pi}\frac{1}{2\pi}\int_{-\pi}^{\pi}\int_{-\pi}^{\pi} \cos([\Omega_1 m+\Omega_2 n]) \ud \Omega_1 \ud \Omega_2, \label{FDM_eq24}
\end{equation}
because imaginary part of \eqref{FDM_eq23} is zero, i.e. $\frac{1}{2\pi}\frac{1}{2\pi}\int_{-\pi}^{\pi}\int_{-\pi}^{\pi} \sin([\Omega_1 m+\Omega_2 n]) \ud \Omega_1 \ud \Omega_2=0$.      

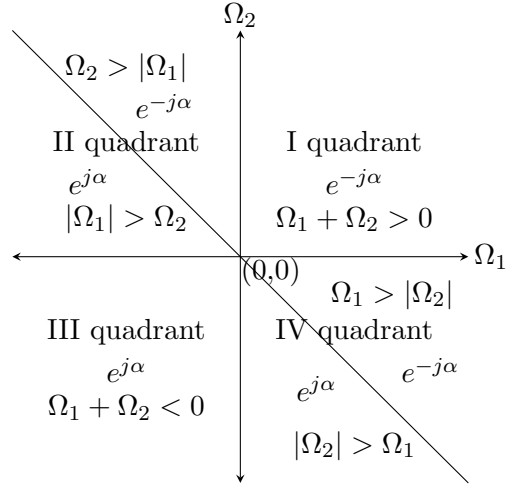
\begin{figure}[!t]
	\begin{tikzpicture}
	
	\draw [<->] (3,3) -- (3,-3);
	\node at (3,3.2) {$\Omega_2$};
	
	\draw [<->] (0,0) -- (6,0);
	\node at (6.3,0) {$\Omega_1$};
	
	%\draw [green, thick, domain=3:-3] plot (\x, {-\x}) node[right] ; 
	\draw [-] (0,3) -- (6,-3);
	
	\node at (3.4,-.2) {(0,0)};
	%\node at (3,0) {II};
	
	\node at (4.5,1.5) {I quadrant};
	\node at (4.5,1) {$e^{-j\alpha}$};
	\node at (4.5,0.5) {$\Omega_1+\Omega_2>0$};

	\node at (4.5,-1) {IV quadrant};
	\node at (5,-.5) {$\Omega_1>|\Omega_2|$};
	\node at (5.5,-1.5) {$e^{-j\alpha}$};
	\node at (4,-1.75) {$e^{j\alpha}$};
	
	\node at (4.5,-2.5) {$|\Omega_2|>\Omega_1$};

	\node at (1.5,1.5) {II quadrant};
	\node at (1.5,2.5) {$\Omega_2>|\Omega_1|$};
	\node at (1.5,0.5) {$|\Omega_1|>\Omega_2$};
	
	\node at (2,2) {$e^{-j\alpha}$};
	\node at (1,1) {$e^{j\alpha}$};

	\node at (1.5,-1) {III quadrant};
	\node at (1.5,-1.5) {$e^{j\alpha}$};
	\node at (1.5,-2) {$\Omega_1+\Omega_2<0$};
	
	\end{tikzpicture}
	\caption{Phase transform regions with four quadrants (I,II,III and IV) in Fourier domain, II quadrant is divided in two parts (a) $\Omega_2>|\Omega_1|$ and (b) $|\Omega_1|>\Omega_2$; IV quadrant is divided in two parts (a) $\Omega_1>|\Omega_2|$ and (b) $|\Omega_2|>\Omega_1$; complete region can be divided in two parts by line $\Omega_1+\Omega_2=0$.}
	\label{fig:Quadrants}
\end{figure}

Now, from Observation \ref{obj1} which is based on the Bedrosian theorem, we conclude that 2D Hilbert transform (2D-HT) as a true extension of 1D-HT can be derived by considering the regions of 2D-DTFT as shown in Figure \ref{fig:Quadrants}, thus we hereby define 2D-PT kernel in frequency domain as
\begin{equation}
H(\Omega_1,\Omega_2,\alpha) =
\begin{cases}
e^{-j\alpha},  \quad \Omega_1+\Omega_2\ge 0, \\
%\cos(\alpha), \quad f=0, \\
e^{j\alpha},  \quad \Omega_1+\Omega_2\le 0.
\end{cases}\label{2DPTK}
\end{equation}
Using \eqref{2DPTK}, we write \eqref{FDM_eq24} as 
\begin{equation}
\delta[m,n]=\frac{1}{\pi}\frac{1}{2\pi}\int_{-\pi}^{\pi}\left[\int_{-\Omega_2}^{\pi} \cos([\Omega_1 m+\Omega_2 n]) \ud \Omega_1\right] \ud \Omega_2, \label{dlt1}
\end{equation}
and define HT of \eqref{dlt1} as
\begin{equation}
h[m,n]=\delta[m,n,\pi/2]=\frac{1}{\pi}\frac{1}{2\pi}\int_{-\pi}^{\pi}\left[\int_{-\Omega_2}^{\pi} \sin([\Omega_1 m+\Omega_2 n]) \ud \Omega_1\right] \ud \Omega_2, \label{dlt2}
\end{equation}
and obtain 
\begin{equation}
h[m,n] =
\begin{cases}
0, \quad m = 0, n=0,\\
\frac{\delta[n-m]}{\pi m},  \quad m=n, m\ne 0, n\ne 0,\\
\frac{-\cos(\pi m)}{\pi m}, \quad m\ne 0, n= 0,\\
\frac{-\cos(\pi n)}{\pi n}, \quad m= 0, n\ne 0,\\
0, \quad m\ne n \ne 0,
\end{cases}\label{dlt3}
\end{equation}
%\begin{equation}
%h[m,n] =
%\begin{cases}
%\frac{\delta[n-m]}{\pi m},  \quad m=n, m\ne 0, n\ne 0,\\
%0, \quad \text{otherwise},
%\end{cases}\label{dlt3}
%\end{equation}
which implies kernel of 2D constant PT can be written 
\begin{equation}
\delta[m,n,\alpha]=\cos(\alpha) \delta[m,n]+\sin(\alpha)\delta[m,n,\pi/2], \label{zn1}
\end{equation}
and thus
\begin{equation}
g[m,n,\alpha]=\cos(\alpha) g[m,n]+\sin(\alpha)g[m,n,\pi/2], \label{zn2}
\end{equation}
where $\delta[m,n,\alpha]=\frac{1}{\pi}\frac{1}{2\pi}\int_{-\pi}^{\pi}\left[\int_{-\Omega_2}^{\pi} \cos([\Omega_1 m+\Omega_2 n]-\alpha) \ud \Omega_1\right] \ud \Omega_2$  is 2D-PT of unit impulse sequence $\delta[m,n]$; $g[m,n,\alpha]=\delta[m,n,\pi/2]*g[m,n]$ is 2D-PT of a signal $g[m,n]$ and \eqref{zn2} is 2D-HT of a signal $g[m,n]$ when $\alpha=\pi/2$. Using \eqref{FDM_eq22} and above discussion, we obtain 2D analytic signal (2D-AS) as
\begin{equation}
z[m,n]=\frac{1}{2\pi}\frac{1}{\pi}\int_{-\pi}^{\pi}\int_{-\Omega_2}^{\pi}\left[ G(\Omega_1,\Omega_2) \exp(j[\Omega_1 m+\Omega_2 n]) \ud \Omega_1 \ud\right] \Omega_2. \label{zn3}
\end{equation}
and its PT as
\begin{equation}
z[m,n,\alpha]=z[m,n]e^{-j\alpha}=g[m,n,\alpha]+jg[m,n,\alpha+\pi/2]. \label{zn4}
\end{equation}

Similar to \eqref{phad}, one can obtain kernel of 2D discrete time analytic signal and compute the phase difference between $\delta[m,n]$ and its HT $h[m,n]$ \eqref{dlt3} as
\begin{multline}
\qquad \qquad \quad z_{\delta,h}[m,n]=\delta[m,n]+jh[m,n], \\
\phi_{\delta,h}[m,n] = \tan^{-1}\left(\frac{h[m,n]}{\delta[m,n]}\right)=
\begin{cases}
0, \quad m = 0, n=0,\\
\pi/2,  \quad m=n, m>0, n>0, \\
-\pi/2,   \quad m=n, m<0, n<0,\\ 
-\pi/2, \quad (m> 0 \text{ \& even) or } (m< 0 \text{ \& odd}), n=0,\\
\pi/2, \quad (m> 0 \text{ \& odd) or } (m< 0 \text{ \& even}), n=0,\\
%\pi/2, \quad m< 0 \text{ \& even}, n=0,\\
%-\pi/2, \quad m< 0 \text{ \& odd}, n=0,\\ 
-\pi/2, \quad m= 0, (n> 0 \text{ \& even) or } (n<0 \text{ \& odd}),\\
\pi/2, \quad m= 0, (n> 0 \text{ \& odd) or } (n< 0 \text{ \& even}).\\
%\pi/2, \quad m= 0, n< 0 \text{ \& even},\\
%-\pi/2, \quad m= 0, n<0 \text{ \& odd}.\\
\end{cases}\label{phad1}
\end{multline}

%\section*{Ethics statement}
%This study did not involve any active collection of human data.

%\section*{Data accessibility statement}
%The FDM MATLAB code is publicly available for download at
%\\\url{https://www.researchgate.net/publication/307606777_MATLABCodeOfBreakingTheLimitsRedefiningTheIF}. All data used in this study are publicly available and links are included in the references of paper.

%\section*{Competing interests statement}
%We have no competing interests.
%
%\section*{Authors' Contributions}
%P Singh carried out mathematical analyses and the simulation work, conceived and designed the study, did data analysis, and drafted the manuscript. Author approved and submitted the manuscript for publication.
%
%\section*{Funding}
%There is no funding to support this research.
%
%\section*{Permission to carry out fieldwork}
%No permissions were required prior to conducting this research.
%
%\section*{Acknowledgment}
%This section does not apply.


\begin{thebibliography}{40}
\itemsep0em % to remove space between references

\bibitem{IEEECompt} N. Ahmed, T. Natarajan, K. R. Rao, {Discrete Cosine Transform}, {IEEE Trans. Computers}, 90--93, (1974).
\bibitem{DCTBook} V. Britanak, P. C. Yip, K. R. Rao, Discrete Cosine and Sine Transforms: General properties, Fast algorithms and
Integer Approximations, February 2006.
\bibitem{CT} Cooley J. W., Tukey J. W., An algorithm for the machine calculation of complex Fourier series, Math. Comput., 19, 297--301, 1965, \url{https://doi.org/10.1090/S0025-5718-1965-0178586-1}

\bibitem{FQT} P. Singh, Novel Fourier Quadrature Transforms and Analytic Signal Representations for Nonlinear and Non-stationary Time Series Analysis, {arXiv:1803.11131 [eess.SP]}, (2018). 
\bibitem{rslc8} P. Singh, {Some studies on a generalized Fourier expansion for nonlinear and nonstationary time series analysis}. {PhD thesis}, Department of Electrical Engineering, IIT Delhi, India (2016).
\bibitem{rslc9} P. Singh, S. D. Joshi, R. K. Patney, K. Saha, The Fourier decomposition method for nonlinear and non-stationary time series analysis. {Proc. R. Soc. A} 20160871 (2017). \url{http://dx.doi.org/10.1098/rspa.2016.0871}.
\bibitem{rslc10} P. Singh, {Time-Frequency analysis via the Fourier Representation}. {arXiv:1604.04992 [cs.IT]}, (2016).
\bibitem{rslc11} P. Singh, S. D. Joshi, {Some studies on multidimensional Fourier theory for Hilbert transform, analytic signal and space-time series analysis}. {arXiv:1507.08117 [cs.IT]}, (2015).
\bibitem{rslc13} {P. Singh, S. D. Joshi, R. K. Patney, K. Saha}, Fourier-based Feature Extraction for Classification of EEG Signals Using EEG Rhythms. {Circuits Syst. Signal Process.} 35(10), 3700--3715 (2016).


\bibitem{MDCT1} J. P. Princen, A. W. Johnson, A. B. Bradley, Subband/transform coding using filter bank designs based on time domain aliasing cancellation, IEEE Proc. Intl. Conference on Acoustics, Speech, and Signal Processing (ICASSP), 2161--2164, 1987.
\bibitem{TDAC} J. P. Princen, A. B. Bradley, Analysis/synthesis filter bank design based on time domain aliasing cancellation, IEEE Trans. Acoust. Speech Signal Processing, ASSP-34 (5), 1153--1161, 1986.


\bibitem{SH} S. Haykin, Communication systems, third edition, John Wiley \& Sons (Asia) Singapore, (1995).
\bibitem{DGabor} D. Gabor, Theory of communication, Electrical Engineers-Part III: Journal of the Institution of Radio and Communication Engineering, 93 (26), 429--441, (1946).

\bibitem{IEEETNSRE}	Gupta A., Singh P., Karlekar M., A novel Signal Modeling Approach for Classification of Seizure and Seizure-free EEG Signals, {\em IEEE Transactions on Neural Systems and Rehabilitation Engineering}, 26 (5), 925--935, 2018. DOI: \url{https://doi.org/10.1109/TNSRE.2018.2818123}.



\bibitem{PSBL} P. Singh, {Breaking the Limits: Redefining the Instantaneous Frequency}, Circuits Syst Signal Process (2017). {https://doi.org/10.1007/s00034-017-0719-y}.
\bibitem{rslc4} {P. Singh, S. D. Joshi, R. K. Patney, K. Saha}, {The Hilbert spectrum and the Energy Preserving Empirical Mode Decomposition}. {arXiv:1504.04104 [cs.IT]}, (2015).
\bibitem{rslc5} {P. Singh, S. D. Joshi, R. K. Patney, K. Saha}, {Some studies on nonpolynomial interpolation and error analysis}. {Appl. Math. Comput.} 244, 809--821 (2014).
\bibitem{rslc6} {P. Singh, P. K. Srivastava, R. K. Patney, S.D. Joshi, K. Saha}, {Nonpolynomial spline based empirical mode decomposition}. {Signal Processing and Communication (ICSC), 2013 International Conference on}, 435--440, (2013).
\bibitem{rslc7} {P. Singh, S. D. Joshi, R. K. Patney, K. Saha}, {The Linearly Independent Non Orthogonal yet Energy Preserving (LINOEP) vectors}. {arXiv:1409.5710 [math.NA]}, (2014).
\bibitem{rslc12}  P. Singh, {LINOEP vectors, spiral of Theodorus, and nonlinear time-invariant system models of mode decomposition}. {arXiv:1509.08667 [cs.IT]}, (2015).
\bibitem{SSandoval} S. Sandoval, P. L. De Leon, Theory of the Hilbert Spectrum, arXiv:1504.07554 [math.CV], 2015.

\bibitem{th19} J. Carson, T. Fry, Variable frequency electric circuit theory with application to the theory of frequency modulation. {Bell System Tech. J.}, 16, 513--540 (1937).
\bibitem{LCohen} L. Cohen, Time-Frequency Analysis, Prentice Hall (1995)
\bibitem{Gabor} D. Gabor, Theory of communication, Proc. IEE. 93(III), 429--457 (1946)


%\bibitem{JVille} J. Ville, Theorie et application de la notion de signal analytic, Cables et Transmissions, 2A(1), 61--74, Paris, France,
%1948. Translation by I. Selin, Theory and applications of the notion of complex signal, Report T-92, RAND Corporation, Santa Monica, CA.
\bibitem{Shekel} J. Shekel, `Instantaneous' frequency, Proc. IRE 41 (4), 548--548, (1953).
\bibitem{DEVakman} D. E. Vakman, On the definition of concepts of amplitude, phase and instantaneous frequency, Radio Eng. and Electron. Phys. 17, 754--759, (1972).
\bibitem{th4} B. Boashash, Estimating and interpreting the instantaneous frequency of a signal--Part 1: Fundamentals. {Proc. IEEE} 80(4), 520--538 (1992).
\bibitem{th411} B. Boashash, Estimating and interpreting the instantaneous frequency of a signal--Part 2: Algorithms and Applications. {Proc. IEEE} 80(4), 540--568 (1992).
%\bibitem{LCohen} L. Cohen, Time-Frequency Analysis, Prentice Hall, 1995.
\bibitem{LoTa} P. J. Loughlin, B. Tacer, Comments on the Interpretation of Instantaneous Frequency. {IEEE Signal Process. Lett.} 4(5), 123--125 (1997).
%\bibitem{rslc1} {N.E. Huang, Z. Shen, S. Long, M. Wu, H. Shih, Q. Zheng, N. Yen, C. Tung, H. Liu}, {The empirical mode decomposition and Hilbert spectrum for non-linear and non-stationary time series analysis}. {Proc. R. Soc. A}, 454, 903--995, (1998).
\bibitem{blBB} B. Boashash, Time Frequency Signal Analysis and Processing: A Comprehensive Reference, Elsevier, Boston (2003).
%\bibitem{SLRM} S.L. Marple, Computing the Discrete-Time Analytic Signal via FFT, {IEEE Transactions on Signal Processing}, 47 (9), 2600--2603, (1999).

%\bibitem{rslc71} I. Daubechies, J. Lu, H.T. Wu, Synchrosqueezed Wavelet Transforms: an Empirical Mode Decomposition-like Tool. {Appl. Comput. Harmon. Anal.}, 30, 243--261, (2011).
\bibitem{th20} B. Van der Pol, The fundamental principles of frequency modulation, {Proc. IEE}. 93(111), 153--158 (1946)
%\bibitem{EQ33} [Online:] \url{http://www.vibrationdata.com/elcentro.htm}

%\bibitem{GWPRL} B.P. Abbott et al., Observation of Gravitational Waves from a Binary Black Hole Merger. {Phys. Rev. Lett.} PRL 116, 061102 (1--16) (2016).
%\bibitem{GWdata} \url{https://losc.ligo.org/events/GW150914/}
\bibitem{PHT} S. Purves, Phase and the Hilbert transform, The Leading Edge 33, 10 (2014), pp. 1164--1166, \url{https://doi.org/10.1190/tle33101164.1}

\bibitem{Bedrosian} E. Bedrosian, A product theorem for Hilbert transforms, Proc. IEEE 51 (1963), 868--869.
\bibitem{JLB} J. L. Brown, A Hilbert transform product theorem, Proc. IEEE 74 (1986), 520--521.

\bibitem{awt1} J. M. Lilly, S. C. Olhede, On the Analytic Wavelet Transform, IEEE Transactions on Information Theory, vol. 56, no. 8, pp. 4136--4156, 2010.
\bibitem{awt2} J. M. Lilly, S. C. Olhede, IEEE Transactions on Signal Processing, vol., 60, issue 11, pp. 6036--6041, 2012.
\bibitem{awt3} M. Holschneider, Wavelets: An Analysis Tool. Oxford: Oxford Univ. Press, 1998.
\bibitem{Matlab1} \url{https://www.mathworks.com/help/wavelet/gs/inverse-continuous-wavelet-transform.html}

\bibitem{asbw1} J Gao, X. Dong, W.B. Wang, Y. Li, and C. Pan, Instantaneous Parameters Extraction via Wavelet Transform, IEEE Transactions on Geoscience and Remote Sensing, vol. 37, issue 2, pp. 867--870, 1999.
\bibitem{asbw2} M. Taner, F. Koehler, and R. E. Sheriff, Complex seismic trace analysis, Geophysics, vol. 44, pp. 1041--1063, 1979.
\bibitem{asbw3} G. Jinghuai, W. Wenbing, and Z. Guangming, Wavelet transform and instantaneous attributes analysis of a signal, Acta Geophys. Sinica, vol. 40, pp. 821--832, 1997.

\end{thebibliography}
\end{document}